\documentclass[10pt,preprint]{aastex}
\usepackage{psboxit}

\PScommands

\newcommand{\Sersic}{S\'ersic}

\def\simless{\mathbin{\lower 3pt\hbox
    {$\,\rlap{\raise 5pt\hbox{$\char'074$}}\mathchar"7218\,$}}} 
\def\simgreat{\mathbin{\lower 3pt\hbox
    {$\,\rlap{\raise 5pt\hbox{$\char'076$}}\mathchar"7218\,$}}} 

\newcounter{thefigs}
\newcommand{\fignum}{\arabic{thefigs}}

\newcounter{thetabs}
\newcommand{\tabnum}{\arabic{thetabs}}

\newcounter{address}

\shortauthors{Blanton {\it et al.} (2006)}
\shorttitle{Improved background subtraction for SDSS}

\begin{document}

\title{Improved background subtraction for the Sloan Digital Sky
  Survey images}

\author{
Michael R. Blanton\altaffilmark{\ref{NYU}},
Eyal Kazin\altaffilmark{\ref{NYU}},
Demitri Muna\altaffilmark{\ref{NYU}},
Benjamin A. Weaver\altaffilmark{\ref{NYU}},
Adrian Price-Whelan\altaffilmark{\ref{NYU}} 
}

\setcounter{address}{1}
\altaffiltext{\theaddress}{
    \stepcounter{address}
    Center for Cosmology and Particle Physics, Department of Physics, New
    York University, 4 Washington Place, New
    York, NY 10003
    \label{NYU}}

\begin{abstract}
We describe a procedure for background subtracting Sloan Digital Sky
Survey (SDSS) imaging that improves the resulting detection and
photometry of large galaxies on the sky.  Within each SDSS drift scan
run, we mask out detected sources and then fit a smooth function to
the variation of the sky background.  This procedure has been applied
to all SDSS-III Data Release 8 images, and the results are available
as part of that data set.  We have tested the effect of our background
subtraction on the photometry of large galaxies by inserting fake
galaxies into the raw pixels, reanalyzing the data, and measuring them
after background subtraction.  Our technique results in no
size-dependent bias in galaxy fluxes up to half-light radii $r_{50}
\sim 100$ arcsec; in contrast, for galaxies of that size the standard
SDSS photometric catalog underestimates fluxes by about 1.5 mag. Our
results represent a substantial improvement over the standard SDSS
catalog results and should form the basis of any analysis of nearby
galaxies using the SDSS imaging data.
\end{abstract}

\section{Why improve the sky subtraction?}
\label{sec:intro}

The Sloan Digital Sky Survey (SDSS; \citealt{york00a}) is the largest
existing survey of the sky to date; the SDSS-III program has now
released Data Release 8 (DR8; \citealt{eisenstein11a, aihara11a}).
This release provides reliable, well-calibrated catalogs down to
$r\sim 22.5$ over 14,500 square degrees of sky in five bands ($u$,
$g$, $r$, $i$ and $z$).  In so doing, it has provided images of a
large fraction of the known bright galaxies on the sky --- the Messier
and NGC objects, as well as lower surface brightness nearby
galaxies. However, no reliable catalog of the properties of these
galaxies yet exists, primarily because no standard background
subtraction procedure has been implemented that properly treats the
largest, brightest galaxies in the sky. The purpose of this paper is
to provide just such a background procedure, which has now been run
over the entire SDSS imaging dataset, and which is being released as
part of DR8.

Background subtraction of astronomical images is probably a formally
impossible task.  The ideal treatment of the data gathered by a
detector would be to explain (at reasonable $\chi^2$) the counts in
each pixel, using a physical model of the sky brightness and its
gradients, the telescopic optics and scattered light properties, the
astronomical and other sources of light, and the sensitivity, noise
and other properties of the detector.
However, this task is impractical (and likely intractable as well),
given the detail needed in such a model, the time variability of
conditions and instruments, and the probable necessity of treating all
the data simultaneously. For this reason, most practical applications
of background subtraction tend to focus on simple approximations that
are tractable as well as close to correct for the problems of
interest.

For the SDSS, the standard photometric pipeline, named {\tt photo},
takes just such a practical and accurate approach
(\citealt{lupton01a}). In the catalog released in Data Release 7 (DR7;
\citealt{abazajian09a}), {\tt photo} calculates the median-smoothed
background on a scale of $100\times100$ arcsec and subtracts that from
the image before object detection and measurement.  In the catalog
release in DR8, a more sophisticated approach is used, which first
models the brightest galaxies in each field so that they do not
contaminate the sky estimate as much (see \citealt{aihara11a} for a
more complete description).

For faint point source photometry the above approach is highly
accurate (at least, as long as confusion is not a problem). In almost
all cases, any diffuse light is subtracted away (whatever its source)
while the point source itself is left untouched. For these sources, it
is not necessary to fully model the background --- just to separate
point sources from any diffuse sources. The resulting fluxes of stars
are generally repeatable at well below the percent level; that is, as
well as can be expected given the vagaries of determining the overall
calibration of the data set (\citealt{padmanabhan07a}).

However, for galaxies the approach that {\tt photo} uses is more
troublesome.  Whereas it is appropriate for galaxies of small enough
angular extent, for larger galaxies it can be inappropriate.
Typically, this method of sky subtraction causes an underestimate in
their flux, size, and concentration (\citealt{west05a, blanton05b,
hyde09a, west10a}). This problem becomes particularly evident for the
nearby brightest cluster galaxies (\citealt{bernardi07a,
lauer07a}). While the DR8 algorithms represent a modest improvement in
the sky subtraction, the bulk of the problem remains.

In \S\ref{sec:model} below we present a different method for sky
subtraction that is more accurate for large galaxies (while retaining
most of the accuracy for point sources). Our method begins with the
estimate of the sky background from a processed SDSS imaging run.  It
builds masks around all bright detected sources, and any known sources
close to but outside the edge of the imaging run. Then it fits a
smooth spline to the unmasked data, applying appropriate constraints
to regularize the problem in the presence of the heavy masking.

In \S\ref{sec:mosaic} we describe how we take sky subtracted SDSS
fields and mosaic them into single image, for the purposes of
analyzing large galaxies.  


In \S\ref{sec:endtoend}, we test the resulting photometry of these
images in three ways.  First, we compare aperture photometry of point
sources to the standard SDSS aperture photometry, to quantify how much
degradation our mosaicking procedure and sky subtraction have
introduced.
Second, we insert fake bright galaxies into the raw data and reanalyze
the data completely.  We then compare our background subtracted
mosaicked images to the original fake data to evaluate the fidelity of
our procedure. We also evaluate the performance of two versions of the
standard SDSS pipelines (the ones used in SDSS DR7 and DR8).
Third, we compare to the only other publicly available mosaicking and
background subtraction facility for SDSS that we are aware of, the
Montage tool distributed by IRSA (\citealt{berriman03a}); for bright
galaxies we find excellent agreement. 


In \S\ref{sec:summary}, we summarize. The results described here are
available on a field-by-field basis in the SDSS DR8 data
set.\footnote{\tt http://www.sdss3.org/dr8} Tools for returning
mosaics are also available.\footnote{{\tt
    http://data.sdss3.org/mosaics}} Finally, a previous version of the
results, for DR7, is also still available.\footnote{\tt
  http://sdss.physics.nyu.edu/sdss3/mosaic/}

\section{A model fit to the SDSS sky background}
\label{sec:model}

\subsection{The SDSS imaging data}
\label{sec:data}

The SDSS has taken $ugriz$ CCD imaging of $14,555~\mathrm{deg^2}$ of
the sky \citep{york00a,aihara11a}.  Automated software performs all of
the data processing: astrometry \citep{pier03a}; source
identification, deblending and photometry \citep{lupton01a};
photometricity determination \citep{hogg01a}; and calibration
\citep{fukugita96a,smith02a, padmanabhan07b}.

As \citet{gunn05a} describe, the SDSS focal plane has 30 imaging CCDs
evenly spaced (with gaps) across it, six for each of the five filters
($u$, $g$, $r$, $i$ and $z$). Each exposure is taken in drift scan
mode, such that a point in the sky passes through each CCD
sequentially with a gap of 71.72 seconds on average.  This procedure
produces six long rectangular images of the sky, called ``camcols.''
In the standard SDSS pipeline, each camcol is further broken into
multiple ``fields,'' each with a width of 1361 pixels in the scan
direction.

The native pixel scale on the CCDs is approximately 0.396 arcsec. When
we refer to pixels in this paper it is this native scale we will
mean. When we discuss these images, we will refer to the position in
the scan direction as the ``pixel row'' or $y$, and the position
perpendicular to the scan direction as the ``pixel column'' or $x$.

\begin{figure}
\figurenum{\fignum}
\epsscale{0.60}
\plotone{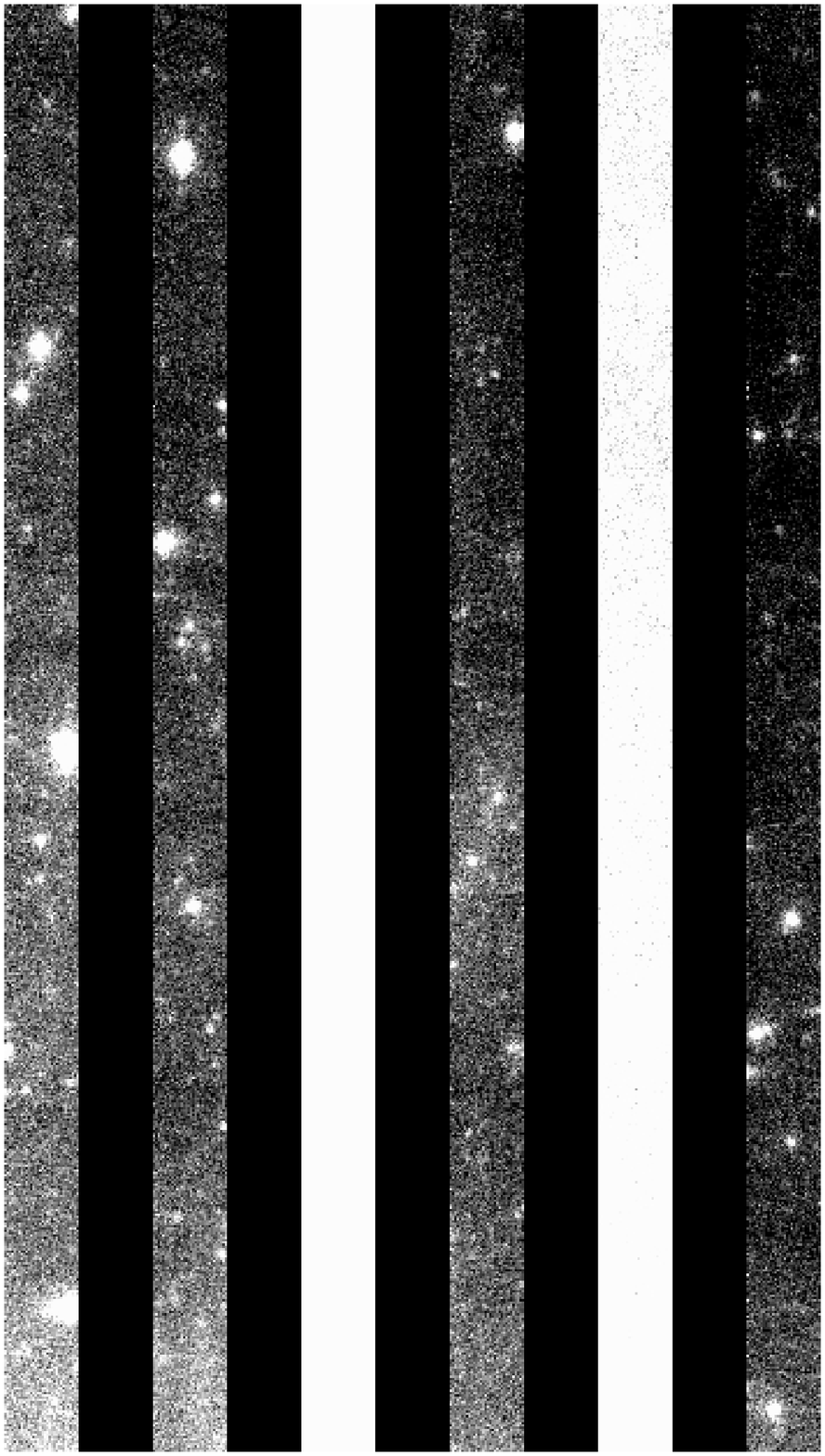}
\caption{\label{fig:rawrun} Data that our sky model is fit to, for
part of an SDSS drift scan run.  The vertical direction is the scan
direction ($y$), the horizontal direction is perpendicular to the scan
direction ($x$). This image is the 8$\times$8 binned flat-fielded SDSS
data. Areas where no objects were detected show the original data.
Areas where objects were detected are replaced by the background sky
estimate plus noise. Each of the six vertical stripes represents a
``camcol'' and the black areas in between are not covered by a
CCD. The units of the image are raw counts, which therefore reflect
the relative gains of the different CCDs. }
\end{figure}

Figure \ref{fig:rawrun} shows this geometry for part of run number
1336 in the $r$-band, showing each camcol and the gaps in between each
one.  It shows an area about 2.3 deg wide representing about 25
minutes of exposure time.  This image is the standard SDSS pipeline's
estimate of the background light used for photometry, binned in
8$\times$8 pixels --- that is, detected objects have been removed from
this image.  In detail, it is equal to the actual counts in areas
where there were no detected objects, and is equal to the background
estimate plus appropriate noise in areas where there were detected
objects. This determination is made before any calibration except for
the flat-field correction, which is only a function of pixel column
because the observations are drift scans.  The units are raw counts;
because of the varying gains among the CCDs two of camcols mostly
saturate in this particular image stretch.

Inspection of Figure \ref{fig:rawrun} reveals that residual light from
bright objects is easily visible in the background light image. This
residual light is due to the SDSS's background light estimation
method, which consists of a median-smoothed image on a scale of
100$\times$100 arcsec.  Under this procedure, large objects have a
substantial amount of their light removed during background
subtraction --- therefore, the background shows the presence of bright
stars.  We will proceed by fitting a much smoother function to this
SDSS sky estimate in order to model the obvious variation with the
background with time but not oversubtract the brightest objects.


\subsection{Masking the data}
\label{sec:mask}

The initial step in our background subtraction is to mask the data
appropriately. We cannot allow the subtraction to be strongly affected
by the presence of bright stars or galaxies, if we want to accurately
measure the fluxes of those objects. We begin by defining an initial
mask around known bright objects.

First, we identify any objects detected in the SDSS catalog with
magnitude $m<15$ in the given filter. We create a mask of size
32$\times$32 native SDSS pixels around the faintest such objects,
growing with decreasing magnitude to 1600$\times$1600 pixels at $m=12$
(and constant with magnitude for brighter objects).  This procedure
masks successfully finds and masks bright stars inside the imaging
run.

However, galaxies within the imaging run are frequently shredded into
fragments by the SDSS deblender, and it is more complex to determine
what area to mask based on the SDSS catalog alone. Therefore, we also
identify any galaxies from the Third Reference Catalog (RC3;
\citealt{devaucouleurs91a}) that land within the imaging run and build
masks around those objects that are 1000$\times$1000 pixels.  This
procedure masks a number of bright galaxies that are large enough to
affect our background subtraction but that our first step above would
not catch.

Finally, the SDSS filters suffer from internal reflection at their
edges, causing stars just outside the frame to scatter light onto the
CCD, creating large-scale features that we do not want included in our
background subtraction model. (A more detailed model would explicitly
subtract these features, that we do not attempt here).  We mask these
reflections by identifying any stars from the Tycho-2 catalog that lie
just outside the frame and masking a rectangular region
1500$\times$900 pixels centered on the star. We use a rectangular
region because the reflection has a larger extent in the pixel column
direction than in the row direction.

Occasionally SDSS fields contain objects sufficiently bright that the
SDSS pipeline fails to reduce them at all.  We mask the entirety of
those fields.

\begin{figure}
\figurenum{\fignum}
\epsscale{0.60}
\plotone{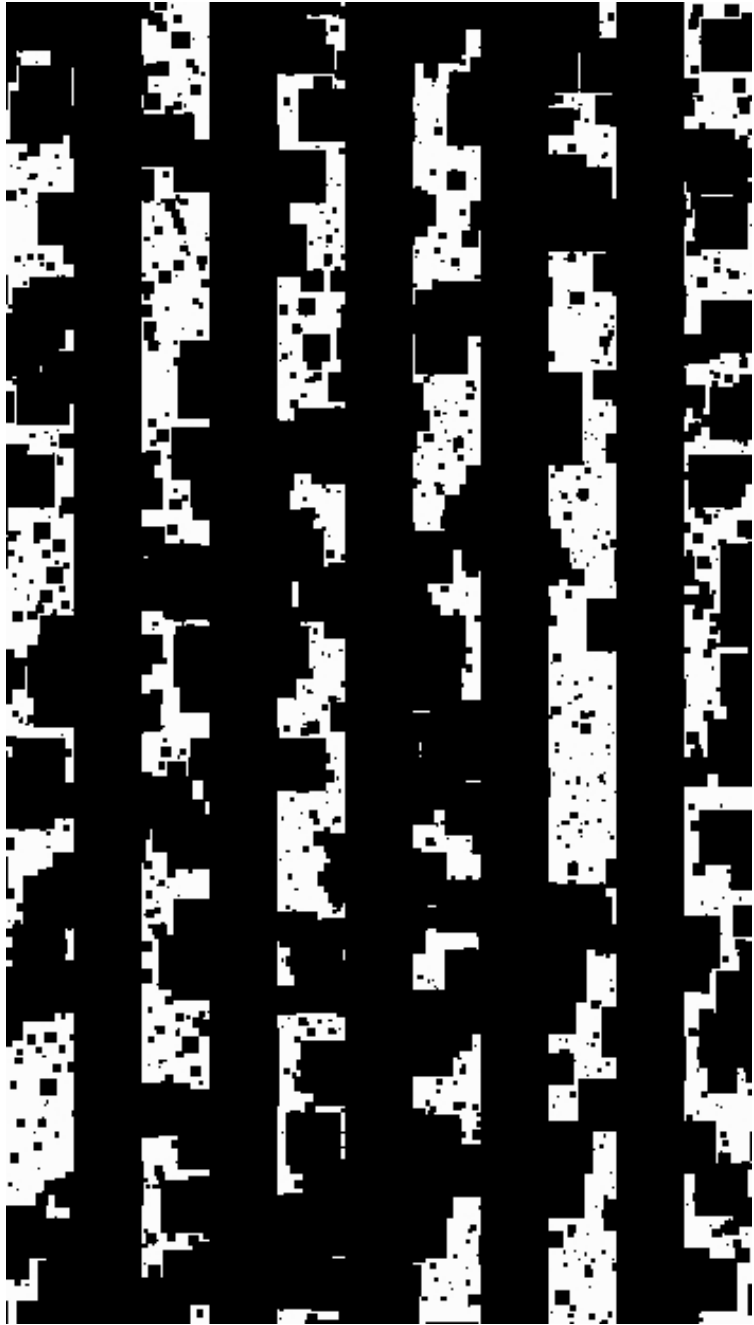} 
\caption{\label{fig:mask} Mask applied to the data, for the same area
of sky as shown in Figure \ref{fig:rawrun}. White areas contribute to
the fit, black areas do not.}
\end{figure}

Figure \ref{fig:mask} shows this initial mask for the run shown in
Figure \ref{fig:rawrun}. White areas are those to be used in our
background model fit while black areas are those to be ignored. The
mask is intentionally very conservative --- we want to minimize the
flux that is incorrectly assigned to the sky background.  As described
in the next section, we also apply further masking to the run during
our iterative fit.

\subsection{Spline model fit}
\label{sec:spline}

In order to fit the data, we first further bin it from 8$\times$8 to
8$\times$680 pixels --- that is, we heavily bin it in the pixel row
direction (i.e. parallel to the scan direction).  We keep the
resolution in the pixel column direction because the sky ends up
tracking some residual flat-fielding errors. When rebinning, we
account for the mask, and track the weight of each 8$\times$680 binned
pixel as the number of non-masked 8$\times$8 pixels it contains.

The spline model that we fit to the final binned data is as follows:
\begin{equation}
\label{skyspline}
    f(\mathrm{camcol}, x, y) = S(\mathrm{camcol}) \times
    Y(\mathrm{camcol}, y) \times X(\mathrm{camcol}, x, y),
\end{equation}
where $x$ and $y$ are the pixel column and row positions.
$S(\mathrm{camcol})$ is an overall scaling factor for each
camcol. $Y(\mathrm{camcol}, y)$ is a second-order b-spline for each
camcol, with break-points spaced approximately once per SDSS field
(1361 pixels). Thus, it expresses the overall variation of the sky
over time during the scan.  $X(\mathrm{camcol}, x, y)$ is a
second-order b-spline in two dimensions, with break-points spaced once
every 8 pixels in the $x$ direction and spaced once every 40 fields in
the $y$ direction.  Thus, it allows for a rapid variation with column
(which accounts for flat-fielding errors) but allows that pattern to
vary slowly over time.

This function is trilinear in the parameters (linear in $S$ and in the
b-spline parameters of $X$ and $Y$); thus an iterative fit is
required.  Our approach here amounts to an approximation to $\chi^2$
minimization of the data with respect to the model.  

First, we determine $S(\mathrm{camcol})$ by taking a median across
each camcol.

Second, we divide each camcol by $S$ and perform $\chi^2$ minimization
to fit $Y$ alone. We quadratically couple the spline parameters
strongly in the row direction to keep the fit smooth. In addition, we
allow a small quadratic coupling between the camcols in order to
interpolate more smoothly over masked data. These quadratic couplings
retain the linearity of the $\chi^2$ fit.

Third, we divide the data by $S$ and $Y$, and finally perform a second
$\chi^2$ minimization, this time for $X$. Again, we allow coupling
between the spline parameters.  In this case, we allow weak coupling
between the parameters in the row direction but very strong coupling
in the column direction, to keep $X$ as smooth as it can be. In the
$\chi^2$ minimizations for $Y$ and $X$, we use the mask weights, but
no inverse variance weighting.

We have a special condition for CCDs that have two amplifiers rather
than one. For these CCDs, half the pixels are read out by one device
and half by another; therefore, we do not couple the pixels across the
divide, and indeed we observe that the gains of the amplifiers do
drift at the fraction of the percent-level over the course of each
run.

Finally, given the resulting model $f(\mathrm{camcol}, x, y)$, we
estimate an rms dispersion $\sigma$ around the fit, and then mask all
regions that are greater than 2$\sigma$ from the fit (in addition to
the initial mask described in \S\ref{sec:mask}). We then iterate the
fit four more times.

\begin{figure}
\figurenum{\fignum}
\epsscale{0.60}
\plotone{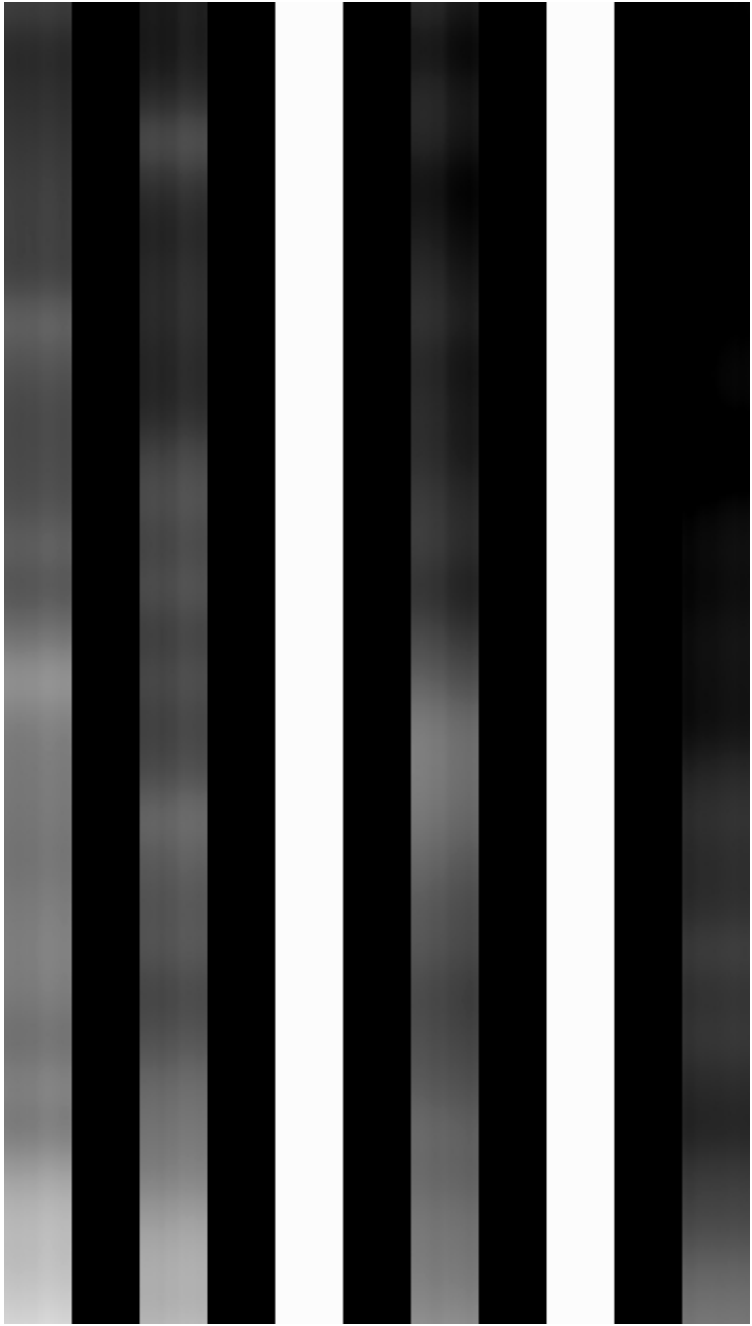} 
\caption{\label{fig:model} Final model sky fit, for the same area of
sky as shown in Figure \ref{fig:rawrun}. This model represents an
evaluation of Equation \ref{skyspline}.}
\end{figure}

An example final fit is shown in Figure \ref{fig:model}. This model is
non-parametric and not physical --- consequently, for diffuse sources
many arcminutes in size, our method will by design oversubtract the
light.  However, \S\ref{sec:simplegal} below shows that it is
sufficiently good to allow accurate measurements of most nearby
galaxies.

\begin{figure}
\figurenum{\fignum}
\epsscale{0.60}
\plotone{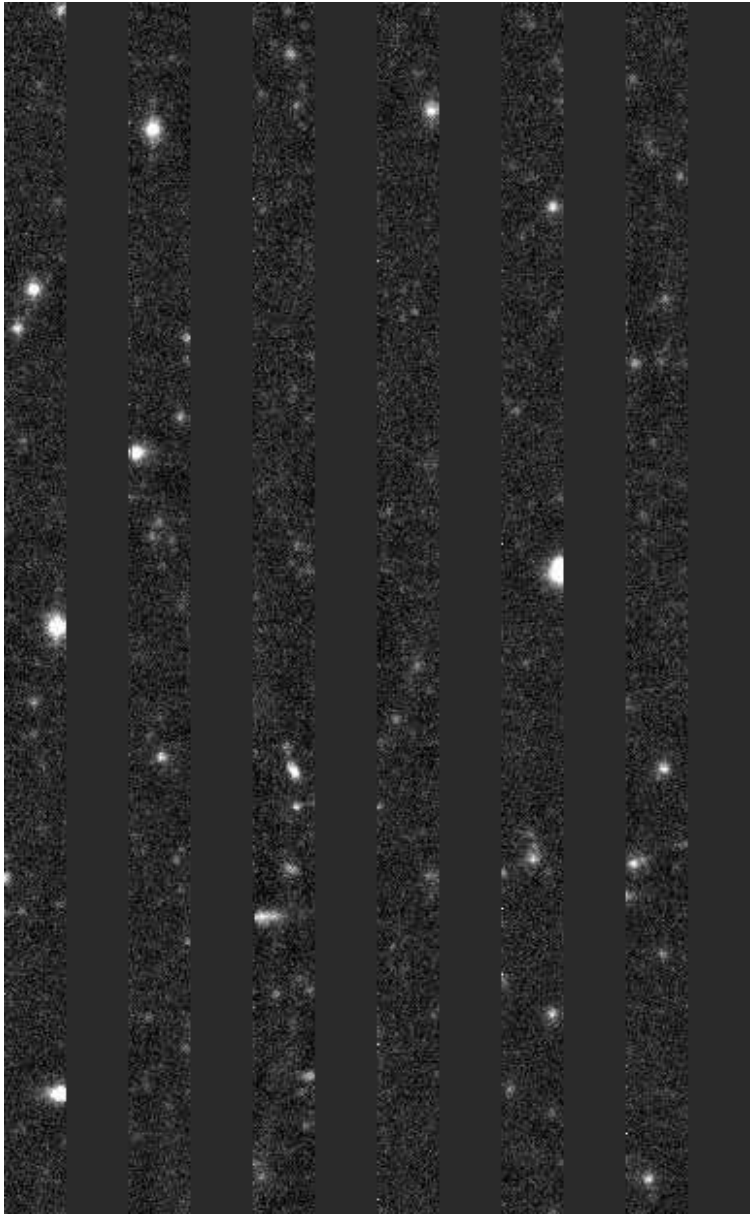} 
\caption{\label{fig:resid} The residuals of the data from the model
(literally Figure \ref{fig:rawrun} minus Figure \ref{fig:model}).}
\end{figure}

Figure \ref{fig:resid} shows the residuals of the data after
subtracting the model sky.  There are remaining discrepancies visible,
which in the next subsection we will show to be at a level that is
fainter than 26 mag arcsec$^{-2}$.

\subsection{Residual tests}
\label{sec:residuals}

We have performed a simple residual test by considering random
unmasked patches that are 13$\times$13 native SDSS pixels in size. We
place these patches in locations that the SDSS photometric pipeline
has identified as ``sky'' objects --- areas with no detected flux in
the pixels.

\begin{figure}
\figurenum{\fignum}
\epsscale{0.60}
\plotone{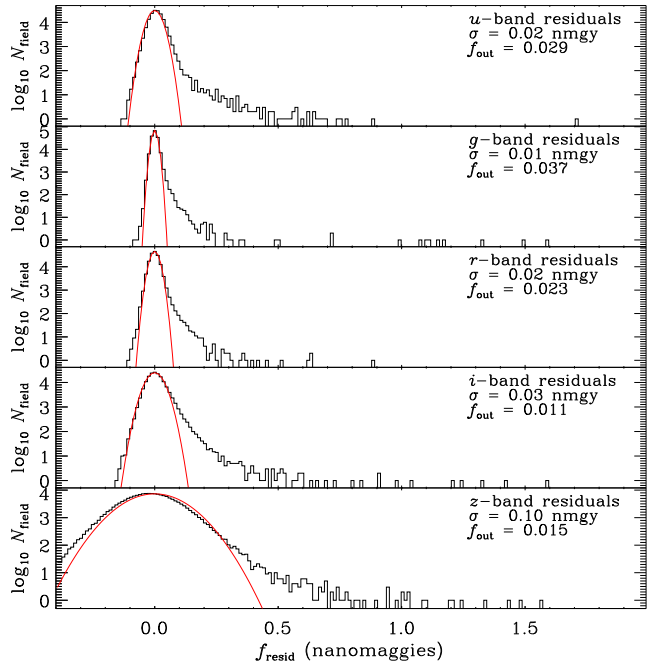} 
\caption{\label{fig:skyqa} Distribution of residuals in each band in
  areas with no detected objects. In each panel, we list the standard
  deviation $\sigma$ of the residuals in nanomaggies per
  arcsec$^{-2}$, and the fraction of areas that are $>3\sigma$
  outliers from the distribution. The smooth line is a Gaussian with a
  standard deviation $\sigma$ for comparison. }
\end{figure}

Figure \ref{fig:skyqa} shows the distribution of the mean fluxes per
square arcsec in such patches across all SDSS runs. The performance in
most areas is excellent, with a standard deviation representing
residuals fainter than $26$ mag arcsec$^{-2}$ in the $r$-band. There
are a handful of outliers, which we find nearly always to be
associated with very extended emission from the interstellar medium or
an extremely bright star. While in regions of very extended emission,
we cannot guarantee that our sky estimate is correct, we do expect
that they will appear as bright outliers in this residual plot.

A smaller contribution to the dispersion in this plot is variable
emission in the night sky lines; this variability mainly affects the
$r$, $i$, and $z$ bands. On some nights, these variations can cause
$\sim 10\%$ fluctuations in the broad-band sky level on time scales of
5 to 10 minutes. Usually our sky subtraction methods remove this
variation well, but in regions which also have very bright objects,
occasionally there are residuals at the few percent level. In
large-scale images, these may show up as red, low surface brightness
streaks.

These residual tests are adequate insofar as they test the performance
of our fitting procedure and help to identify runs with highly
variable backgrounds (astronomical or instrumental).  However, they do
not provide a direct estimate of the effects of sky subtraction errors
on the resulting galaxy photometry --- that is, they do not test
whether we might be oversubtracting light near large sources.  We test
that effect more thoroughly in \S\ref{sec:simplegal}.

\subsection{Distribution of DR8 sky subtracted fields and mosaics}
\label{sec:fields}

The SDSS DR8 data release distributes all of the flat-fielded,
sky subtracted, calibrated fields in each band based on this
procedure. This data consists of approximately 940,000 fields, each
2048$\times$1489 pixels in size, available for download on an
individual field basis (or requesting cutouts) from the SDSS-III
Science Archive Server.\footnote{\tt
  http://data.sdss3.org/}

The data format is FITS. Each file's primary HDU contains a
floating-point image, calibrated in units of nanomaggies per
pixel. The images have had defects such as bad columns and cosmic rays
identified and interpolated over by the {\tt photo} pipeline.

The images are compressed by zeroing out all but the 10 most
significant binary digits in the floating point numbers, and
subsequently using {\tt bzip2} to losslessly compress the result. For
the sky subtracted images, the 10th binary digit is considerably
smaller than the noise in any image, meaning that although overall
this compression is slightly lossy, it has no significant effect on
the results --- even if these images are combined in a stack
(\citealt{pricewhelan10a}).  Subsequent HDUs in each image yield the
sky subtraction and calibration image, so that the sky can be added
back in and the image put back into counts if desired. Full details
of the format and location of the files can be found in the SDSS-III
data model.\footnote{\tt
  http://data.sdss3.org/datamodel/files/BOSS\_PHOTOOBJ/frames/RERUN/RUN/CAMCOL/frame.html}

The Science Archive Server also provides mosaicking tools. They work
slightly differently than the procedure described in
\S\ref{sec:mosaic}.  Instead of using the raw data, they begin with
the corrected frames described here and mosaic the fields together
using the {\tt swarp} package.\footnote{{\tt
    http://www.astromatic.net}} However, they yield equivalent
results.

\section{Mosaicking the SDSS}
\label{sec:mosaic}

\subsection{Generating mosaics}

The primary aim of our sky subtraction is to improve measurements of
large objects on the sky.  Many such objects overlap the edges of SDSS
fields (and some are bigger than or comparable to the size of a single
field).  Thus, to recover their photometry we must mosaic together
multiple fields after the background subtraction. Here we describe our
procedure. For historical reasons, for these mosaics we have not begun
with the calibrated frames described in \S\ref{sec:fields} but with
the raw data.

For each mosaic we specify a desired World Coordinate System header
(WCS; \citealt{greisen02a}). We identify all SDSS fields that overlap
the desired area and pick the minimal set that fully cover that area.

Then we prepare each field for mosaicking.  Starting with the raw SDSS
data, we apply the flat field determined by the ubercalibration
procedure (\citealt{padmanabhan07b}). Next we identify cosmic rays
using a procedure similar to that used by the SDSS pipeline
(\citealt{lupton01a}). We interpolate over saturated pixels and cosmic
rays using simple linear interpolation in the $x$ direction.  We then
subtract our estimate the sky, determined by evaluating the function
described by Equation \ref{skyspline} at each image pixel. In the sky
estimate, we account for any difference in the flat field as
originally used by the SDSS photometric pipeline, and as determined by
ubercalibration. Finally, we apply the photometric calibration to each
field (\citealt{padmanabhan07b}).

To resample each image we evaluate the position of each desired pixel
within each original field. Then we interpolate the original field's
image to the desired set of locations, using the well-known cubic
approximation to the sinc function (\citealt{park83a}). This
resampling method is known to work well for Nyquist-sampled images, a
condition SDSS virtually always satisfies. In practice, we use the
built-in IDL utilities {\tt polywarp} and {\tt poly\_2d} to perform
this resampling.

We evaluate the weighted average of the flux from all the input images
at each output pixel. The weights are unity throughout most of each
input image, but are apodized smoothly to zero at the edges. Thus, the
transition between regions which overlap two fields are relatively
smooth. Any output pixels that have no overlapping input images are
set to zero.

\subsection{Example mosaics}

\begin{figure}
\figurenum{\fignum}
\epsscale{0.70}
\plotone{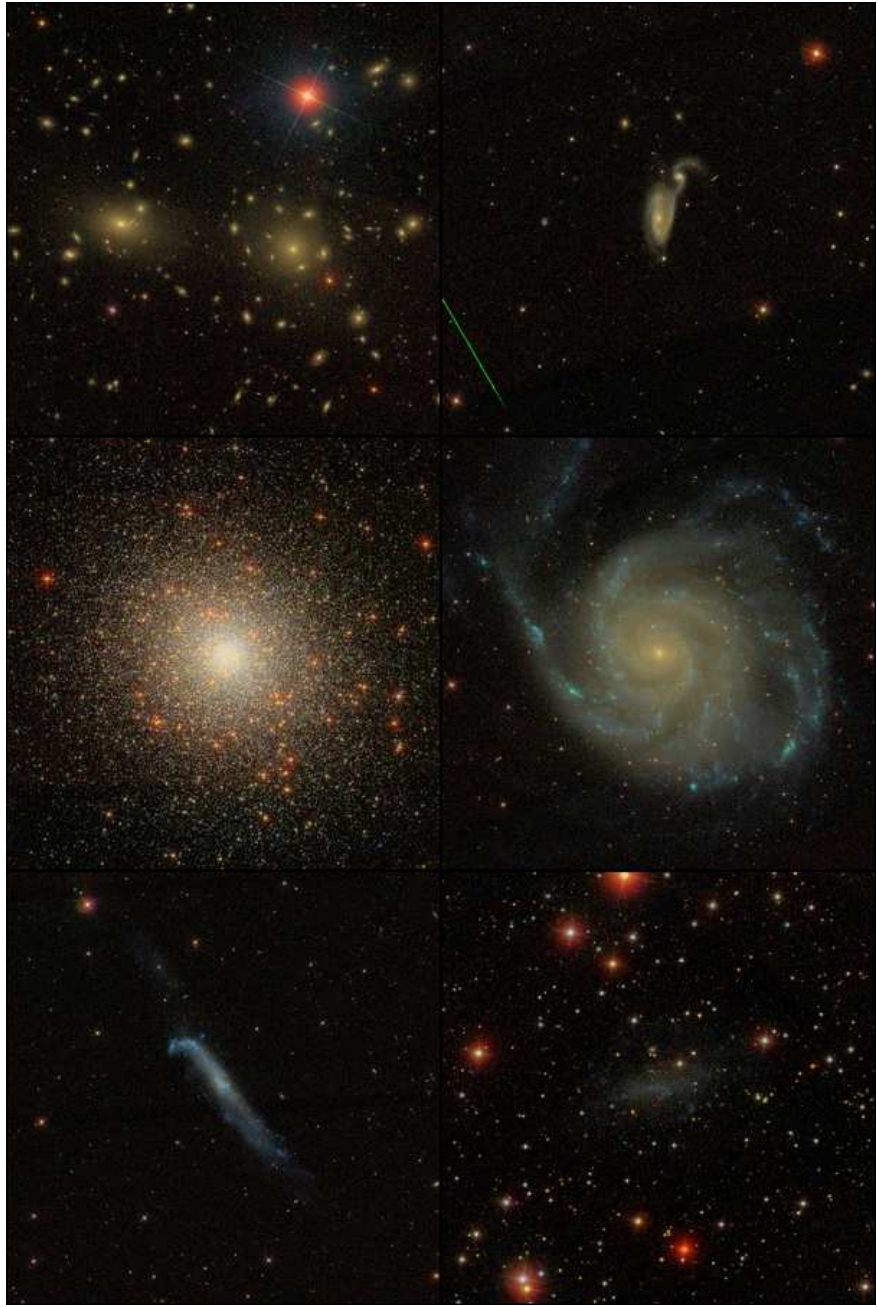} 
\caption{\label{fig:examples} Some example $gri$ composite images made
  with the methods described in this paper. Upper left: Coma Cluster
  ($18'\times 18'$); upper right: NGC 5395 ($18'\times 18'$); middle
  left: Messier 5 ($18'\times18'$); middle right: Messier 101
  ($21'\times 21'$); lower left: NGC 4656 ($24'\times 24'$); lower
  right: UGC 3974 ($12'\times 12'$).}
\end{figure}

Figure \ref{fig:examples} shows several example mosaics created with
this procedure.  We have obviously chosen these partly for dramatic
effect, but also to demonstrate the robustness of the masking to the
presence of large, bright objects of various sorts: globular clusters,
galaxy clusters, and large galaxies. Each of these mosaics is built
from two up to six fields depending on its size (which ranges from 12
to 24 arcmin on a side).

\subsection{Distribution of DR7 mosaics}
\label{sec:distmosaic}

Although our procedures allow us to evaluate any arbitrary mosaic, for
ease of distribution we have created a set of 1$\times$1 deg mosaics
across the entire observed Northern Galactic Cap, or almost 8000
deg$^2$. We have created a web interface that allows users to easily
extract sections of these DR7 mosaics (see footnote in
\S\ref{sec:intro} for URL).

This interface can be used either interactively or
non-interactively. The interactive method allows the user to browse a
Google sky viewer and select the region of interest.  The
non-interactive method is explained in its documentation.  In either
case, the tool returns FITS format images in each desired bandpass,
with correct WCS headers, sky subtracted and calibrated. The units of
the images are nanomaggies per pixel.

Note that we have not yet produced any such mosaics of the area imaged
in SDSS-III's new imaging in the South Galactic Cap. However, as
described in \S\ref{sec:fields} the standard DR8 data release tools do
provide mosaicking tools.

\subsection{Comparison to Montage}

The only other publicly available mosaicking and background
subtraction facility for SDSS that we are aware of is the results of
the Montage package distributed by IRSA
\citep{berriman03a,berriman04a}. We will describe the quantitative
differences between their results and ours in \S\ref{sec:galmontage}.
Here we discuss methodological differences.

The first major difference is in their method for background
subtraction. While in our case we rely on a smooth fit to regions with
no objects, Montage fits a smooth additive term within each run to
minimize the differences between it and other overlapping images.
This procedure can be far superior to ours in regions with many
overlapping images.  In much of the SDSS area the only overlapping
images are at the north and south edges of each field, and how the
Montage algorithm behaves in that regime is a quantitative
question. We compare our results to theirs for large galaxies in
\S\ref{sec:simplegal} below.

The second major difference is in their method for ``resampling'' of
images. They do not perform a normal resampling but instead use a
generalized version of the ``drizzle'' algorithm. This algorithm
computes the fraction of each original pixel that overlaps each output
pixel, and distributes the original pixel's flux accordingly. Relative
to properly resampling a Nyquist-sampled image, such methods do a poor
job at maintaining the original PSF (or even at producing an image
with a well-defined PSF). They are also clearly irreversible
operations even in the noiseless case, demonstrating that information
is lost and the image is consequently degraded. We compare our results
to the Montage results for point sources in \S\ref{sec:psfs} below,
and believe that Montage's are less good at the bright end in those
tests due to their drizzling procedure. However, it is unlikely that
these issues affect our main interest here, which is large galaxies on
the sky.

\section{End-to-end tests of the sky background}
\label{sec:endtoend}

The residual tests of \S\ref{sec:residuals} are acceptable as far as
they go, but are of little use in evaluating the quality of the
photometry of objects in the masked regions.  In order to do that, we
need to compare to ground truth (possible to some extent for point
sources; see \S\ref{sec:psfs}) or simulate the analysis of objects
similar to those we are interested in (necessary for galaxies; see
\S\ref{sec:simplegal} and \S\ref{sec:deblendgal}).

\subsection{PSF photometry}
\label{sec:psfs}

The standard SDSS analysis excels for point sources.  In fact, because
that analysis avoids interpolation, regridding or stacking of the data
we expect that its performance will be superior than anything that our
background subtracted mosaics could produce. Therefore, as a quality
assurance tool to demonstrate how photometric our images are, we
compare aperture photometry of point sources in our mosaics to that
reported by the standard SDSS pipeline. The goal here is to verify
that the overall scale of the mosaicked images is correct and that the
signal-to-noise has not been degraded.

We randomly choose twenty 1 degree square mosaics. Because in this
section we are interested in point source photometry, we perform a
local sky subtraction on the same scales as SDSS uses (100 arcsec
median smoothed).  We of course will not do the same in the analysis
of large galaxies. From the overlapping SDSS catalog, we choose
unsaturated stars with no neighbors within 15 arcsec.  For each we
measure a circular aperture flux within a 7.3 arcsec radius (the
standard aperture used by the ubercalibration pipline).  Then, we
compare this flux to that reported within the same aperture by the
standard SDSS pipeline.

\begin{figure}
\figurenum{\fignum}
\epsscale{1.00}
\plotone{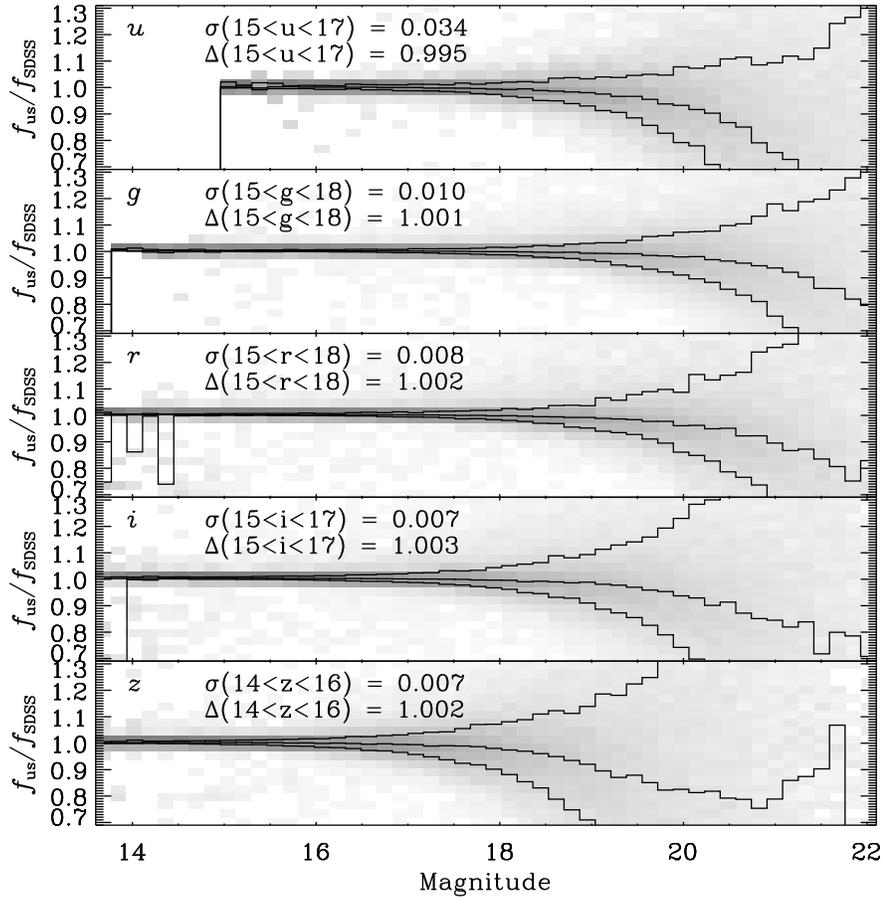}
\caption{\label{fig:sdss_qa_magdiff} Flux ratios between
the 7.3 arcsec aperture flux measured from our mosaic images, and that
reported by the SDSS catalog, for each SDSS band. The lines are the
16\%, 50\% and 84\% quantiles. The grey scale is
proportional to the conditional distribution of the flux ratio on the
magnitude. At bright magnitudes, the differences are small, indicating
that these images have retained their overall photometricity. In each
panel, we list the mean value and dispersion of the flux ratios 
in the
indicated magnitude range. }
\end{figure}

Figure \ref{fig:sdss_qa_magdiff} shows (as a function of magnitude)
the flux ratios between the 7.3 arcsec aperture flux measured from a
random set of twenty of our mosaic images and that reported by the
SDSS catalog for each SDSS band. The overall scale of the magnitudes
is close to correct, with less than a 1\% difference between our
results and SDSS at bright magnitudes.

\begin{figure}
\figurenum{\fignum}
\epsscale{1.00}
\plotone{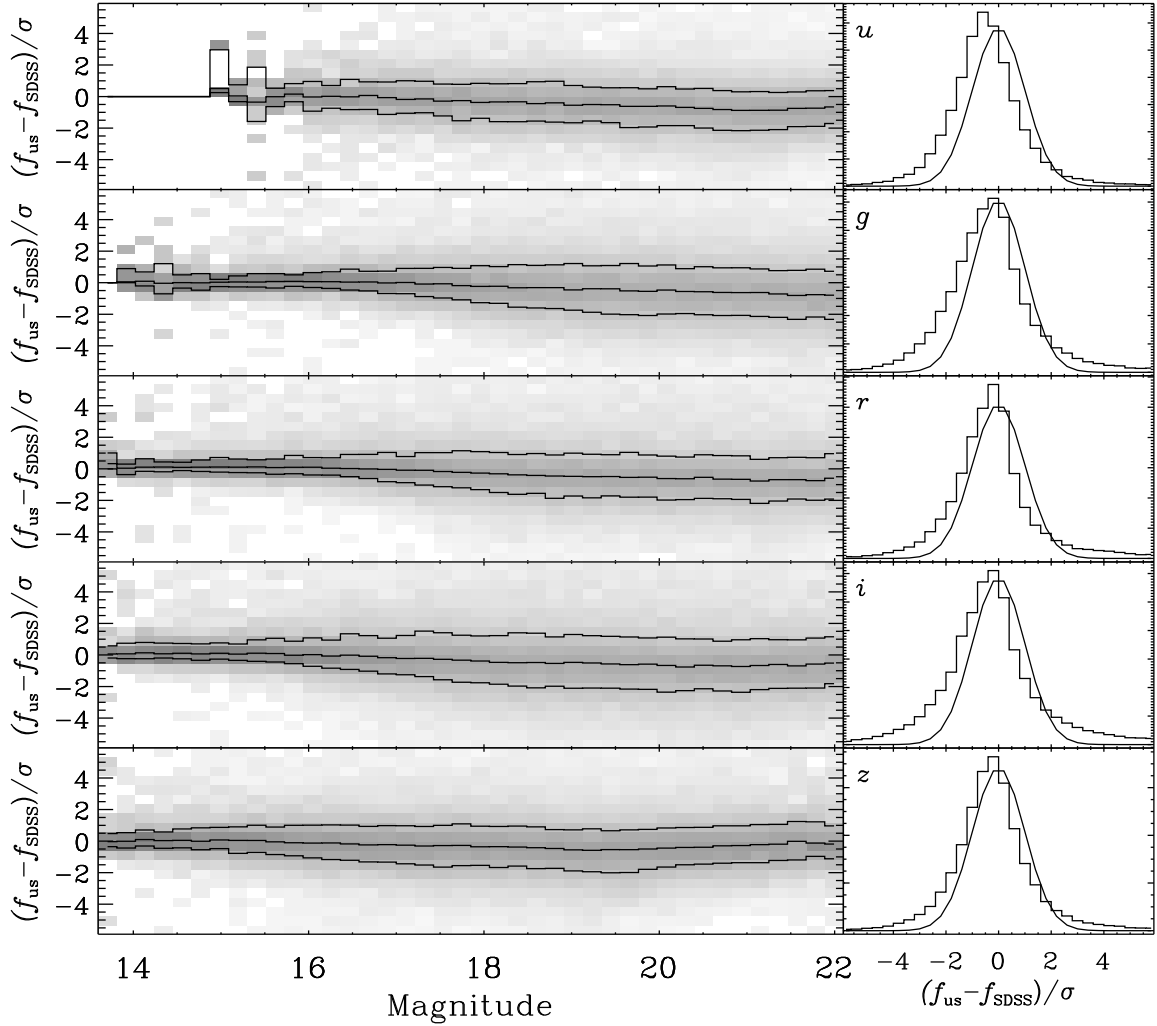}
\caption{\label{fig:sdss_qa_scaled} Similar to Figure
\ref{fig:sdss_qa_magdiff}, but now showing flux differences scaled to
the SDSS aperture flux uncertainty. In the right-hand panels we show
the distribution of each set of differences.  The smooth Gaussian
curve is the expected normal distribution. The flux differences are
not far removed from the normal distribution, indicating that our
images are not much degraded relative to the original images.}
\end{figure}

Figure \ref{fig:sdss_qa_scaled} shows the distribution of flux
differences, scaled to the expected error distribution. The lines are
the 16\%, 50\% and 84\% quantiles. This distribution is close to the
normal distribution, an indication that our mosaicking procedure does
not degrade the image very badly.  Of course it does degrade it to
some degree, an effect particularly noticeable at the faintest
magnitudes, where the median difference becomes almost 0.5$\sigma$ and
the distribution is significantly wider then the normal distribution.

\begin{figure}
\figurenum{\fignum}
\epsscale{1.00}
\plotone{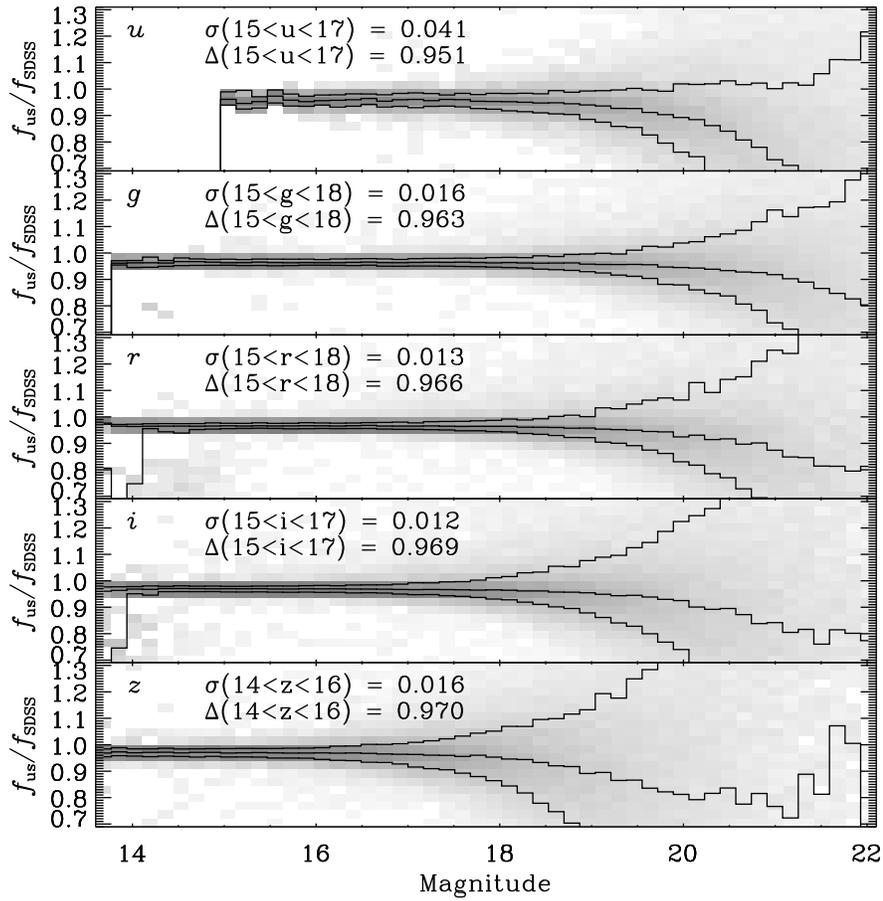}
\caption{\label{fig:montage_qa_magdiff} Similar to Figure
\ref{fig:sdss_qa_magdiff}, but using Montage images instead of our
mosaics. There is an overall scale difference in all bands of about
2--3\% (not a particularly large error if it is uniform). }
\end{figure}

\begin{figure}
\figurenum{\fignum}
\epsscale{1.00}
\plotone{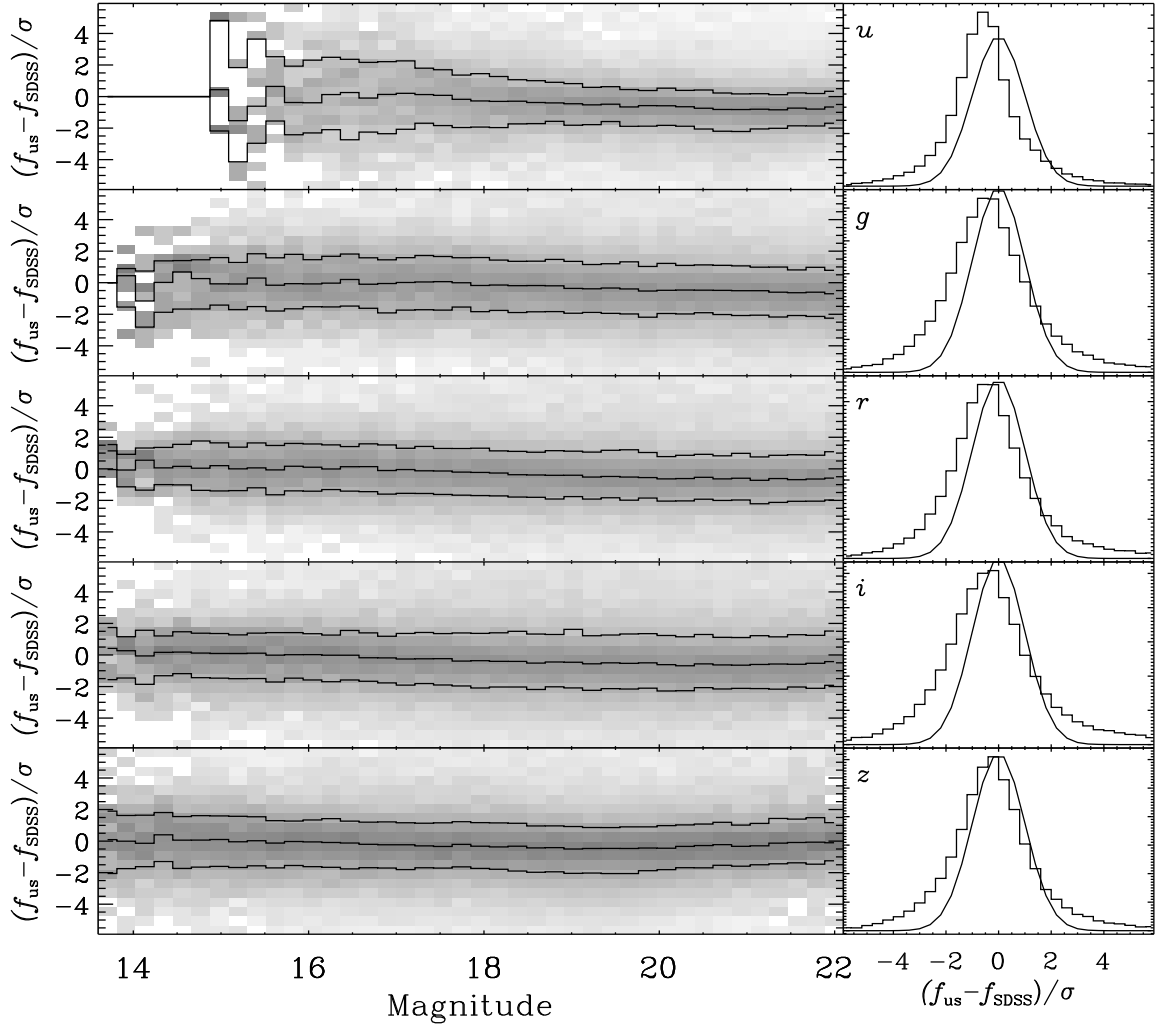}
\caption{\label{fig:montage_qa_scaled} Similar to Figure
\ref{fig:sdss_qa_scaled}, but using Montage images instead of our
mosaics. We account for the overall scale difference found in in each band
before making these plots. The distribution of differences is much
broader than the SDSS uncertainties, indicating that for point sources
Montage images are significantly degraded.}
\end{figure}

We perform the same pair of tests on twenty Montage mosaics of the
same regions; results are shown in Figures
\ref{fig:montage_qa_magdiff} and \ref{fig:montage_qa_scaled}.  The
flux ratios reveal a scale difference between the Montage
images and the SDSS catalog of 3--5\%.
We have not investigated their procedure thoroughly enough to diagnose
the cause of this error.  In any case, as an overall flux error this
problem is relatively minor for our purposes here.

Figure \ref{fig:montage_qa_scaled} shows the distribution of flux
differences between the SDSS catalog photometry, and our aperture
photometry of the Montage mosaics. We have first accounted for the
scale error before considering the flux differences. Having done so,
the error distribution clusters around zero.  At faint magnitudes the
distribution is similar to ours. At bright magnitudes it is
considerably broader, which we suspect is due to the degradation of
the image caused by the drizzling procedure. As we show below, these
errors appear to be less important when considering larger, brighter
objects. In addition, the additional error is extremely small, at the
1\% level.

\subsection{Galaxy photometry: simulated galaxy samples}
\label{sec:fakegal}

In the previous section, we considered the effects of our procedures
on point source photometry to verify that our images did reasonably
well relative to the ``ground truth'' established by the SDSS standard
photometry.  We do not trust the SDSS standard results in the case of
galaxy photometry, --- in fact, it is our goal to improve that
photometry substantially.  Therefore, we must use a different tactic
to test our results.

To do so, we insert fake galaxies into the raw data, using the same
tools developed by \citet{blanton05b} and used by several other
investigators since (\citealt{blanton04b, mandelbaum06a, masjedi06a}).
We distribute fake galaxy images onto a random set of locations on the
sky covered by SDSS fields. For each observation of an object in an
SDSS field in each band, we convert the fake stamps to SDSS raw data
units, convolve with the estimated seeing from the SDSS photometric
pipeline, and add Poisson noise using the estimates of the gain. We
add the resulting image to the real SDSS raw data, including the tiny
effects of nonlinearity in the response and the less tiny flat-field
variation as a function of column on the chip. We then run the SDSS
photometric pipeline {\tt photo} as well as our sky subtraction
procedure.  Around each fake galaxy that we insert, we create a mosaic
using the tools described above.

The fake galaxy sample is based on the low redshift sample from the
New York University Value-Added Galaxy Catalog (\citealt{blanton05a}),
as updated for SDSS Data Release 6 (see, for example,
\citealt{zhu10a}). This sample consists of essentially all galaxies
with $r<17.77$ and $z<0.05$.  For each galaxy, we take its flux,
half-light radius, \Sersic\ index, axis ratio, and $ugriz$ colors from
the SDSS data. We then reassign each galaxy a redshift in the range
$0.0033<z<0.025$ (constrained to be smaller than the actual redshift),
and adjust its size and flux accordingly.  This procedure yields a
realistic distribution of sizes, fluxes, profile shapes and colors,
with an emphasis on the large, nearby galaxies of most interest here.

\begin{figure}
\figurenum{\fignum}
\epsscale{1.00}
\plotone{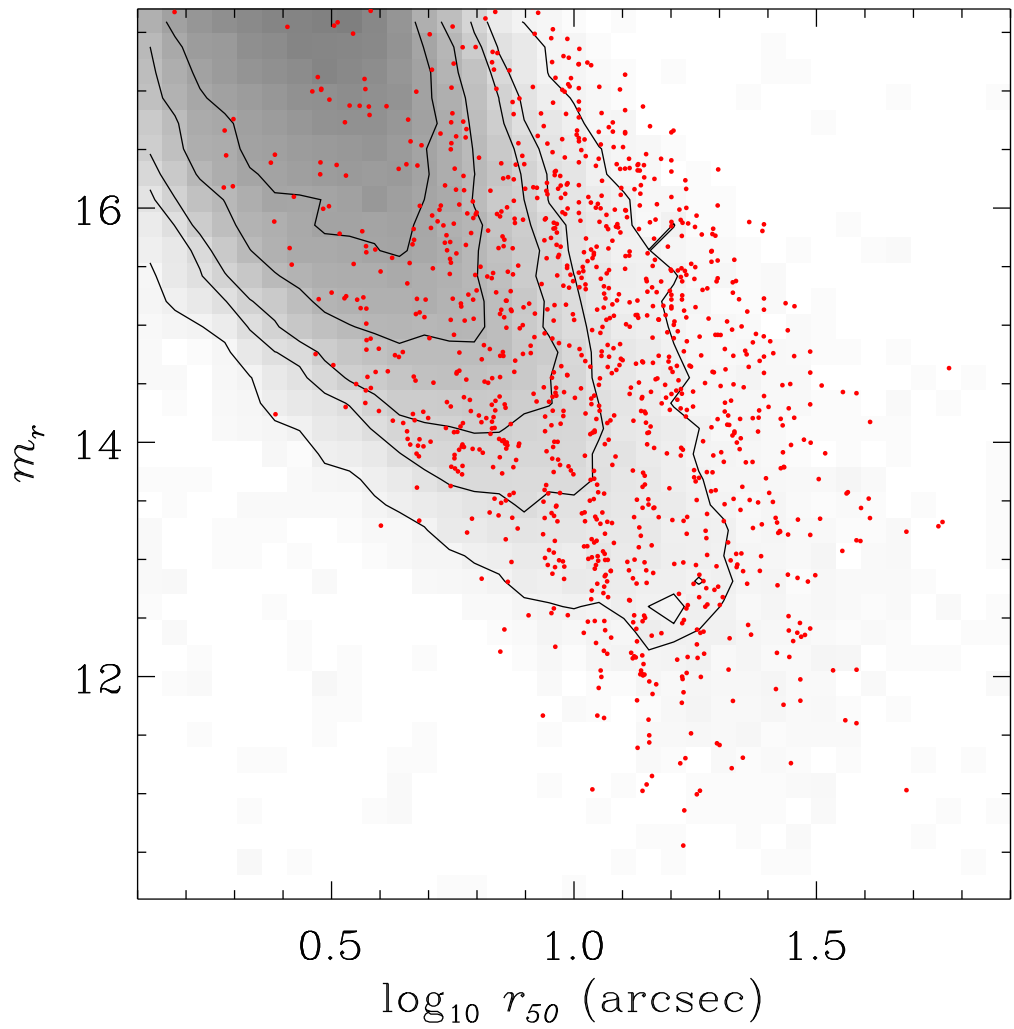}
\caption{\label{fig:fakedist} Distribution of $r$-band measured
  half-light sizes and magnitudes for our fake galaxy sample (points)
  and for real SDSS galaxies (greyscale and contours). The two
  distributions follow each other well, but the fake sample is roughly
  size-limited at $r_{50} \sim 5$ arcsec.}
\end{figure}

There are several different definitions of magnitude in the SDSS
catalogs. Here we will compare with the most robust, the so-called
``cModel'' fluxes, defined in terms of the best fit combination of the
de Vaucouleurs and exponential models as follows:
\begin{equation}
 f_c = f_{\mathrm{exp}} \left(1-\mathtt{fracdeV}\right) +
 f_{\mathrm{deV}}
 \times \mathtt{fracdeV}
\end{equation}
Figure \ref{fig:fakedist} shows the distribution as a function of
measured cModel magnitude and half-light size (from the best-fit model
in each case). The greyscale and contours are from the SDSS DR8
catalog; the points are the fake data (with parameters as measured by
{\tt photo} {\tt v5\_6}). The fake galaxies follow the real
distribution well, though they define a roughly size limited sample
with $r_{50} \simgreat 5$ arcsec.

\subsection{Galaxy photometry: isolating the effect of the background}
\label{sec:simplegal}

Even with background subtracted images, galaxy photometry can be a
complex process, involving deblending the target galaxy from other
sources and a choice of algorithms for estimating the total flux and
other properties.  To test our procedure, we want first to be able to
isolate the effects of the background subtraction and ask whether it
is substantially affecting the results.

To test how well our sky subtraction is doing, we ask directly how
much flux it removes from the aperture surrounding the object. In
particular, we use an aperture with a radius of $r_{95}$, containing
95\% of the light of the galaxy.  We use the fake, sky subtracted
field and calculate the flux in each band in that aperture.  We then
use the real, galaxy-free, sky subtracted field and calculate the flux
in the same aperture.  Of course, in the fake data the sky subtraction
will inevitably subtract off a slightly different (usually larger)
amount of light than does the sky subtraction in the real data,
because of the presence of the large, fake object. The fake aperture
flux minus the real aperture flux is equal to the fake galaxy flux
that remains in the image after sky subtraction --- the measured fake
galaxy flux $f_{\mathrm{meas}}$, to be compared with the original
input fake galaxy flux $f_{\mathrm{true}}$.

In Figure \ref{fig:sky_offsets_ronly}, the squares in the bottom panel
are $\Delta m = -2.5\log_{10}(f_{\mathrm{meas}}/f_{\mathrm{true}})$,
the residual of the measured fake galaxy flux with respect to the
original in the $r$-band as a function of the true galaxy half-light
radius $r_{50}$.  There is no systematic offset of any signficance up
to $r_{50} \sim 100$ arcsec, which indicates a rather large galaxy.
For example, only a couple of dozen galaxies in the DR8 area are that
large. The largest galaxies get to $r_{50}= 10$--$20$ $h^{-1}$ kpc in
size, and such galaxies would have $r_{50} = 100$ arcsec at a distance
of 20--40 $h^{-1}$ Mpc. Thus, we can expect only the largest galaxies
on the sky --- for example, certain Messier objects --- to have any
substantial flux subtracted by this procedure.

Figure \ref{fig:sky_offsets_ugiz} shows the color residuals with
respect to size, in $u-r$, $g-r$, $r-i$, and $i-z$ (measured color
minus true color). Again, the residuals scatter about zero.

The test performed here will typically underestimate the errors of any
actual photometric measurement on the fake galaxies, and defines the
``best possible'' result given our sky subtraction procedure. These
results confirm that our procedure works very well for the photometry
of bright galaxies.

\subsection{Galaxy photometry: full tests of deblended galaxies}
\label{sec:deblendgal}

\begin{figure}
\figurenum{\fignum}
\epsscale{1.00}
\plotone{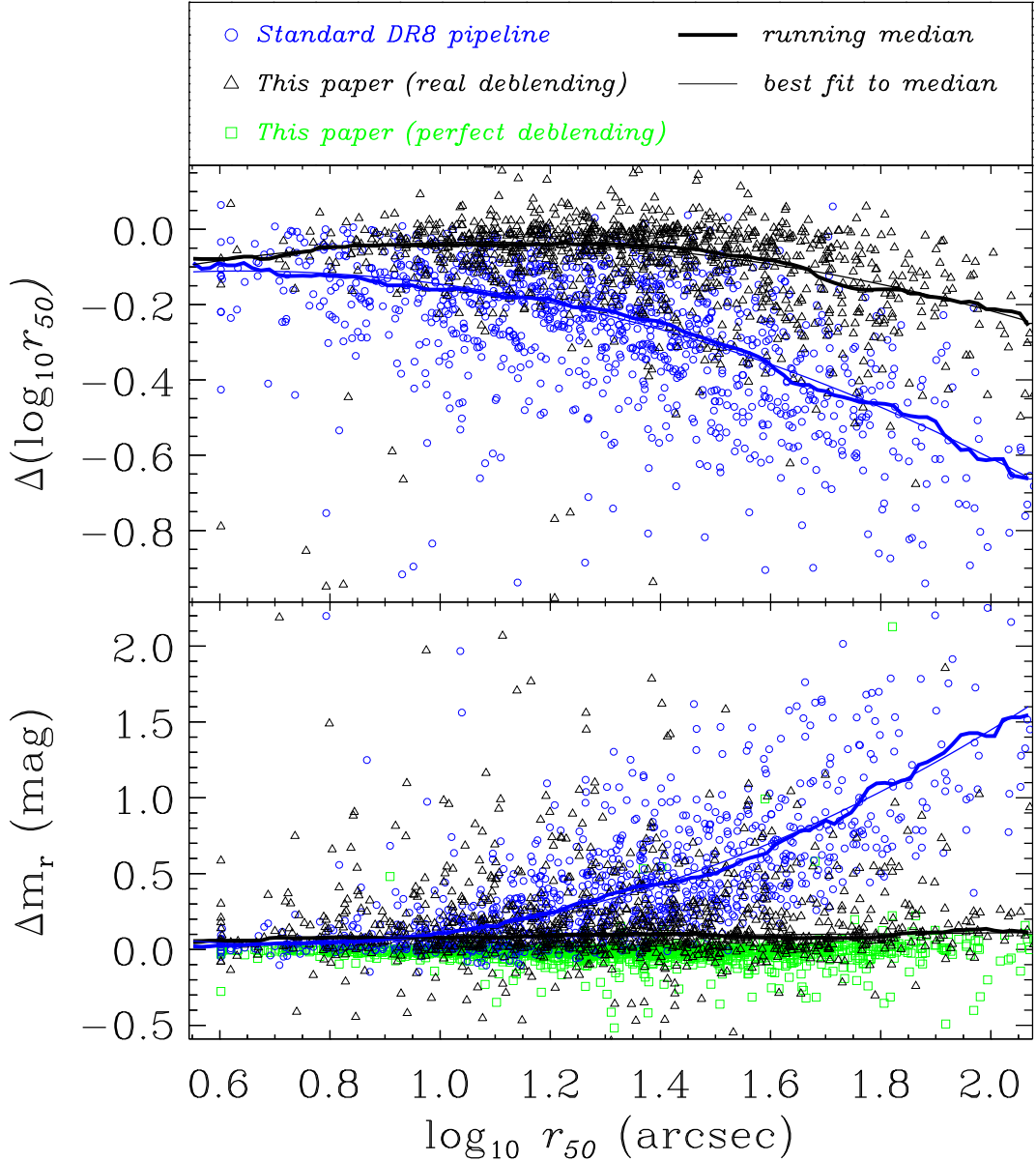}
\caption{\label{fig:sky_offsets_ronly} Errors in half-light radii $r_{50}$
  (top panel) and magnitudes (bottom panel) in the $r$-band, as a
  function of true galaxy $r_{50}$. The open circles show standard
  SDSS catalog photometry from the DR8 version of the pipeline; the
  DR7 version is similar but slightly worse. The triangles show
  photometry and deblending using the methods described in this paper.
  The open squares show the photometry of the fake galaxies assuming
  perfect deblending.  The thick lines are running medians for each
  distribution. The thin lines are the smooth fit given by Equation
  \ref{eq:sky_offsets_model}, with parameters listed in Table
  \ref{table:sky_offsets}.}
\end{figure}

\begin{figure}
\figurenum{\fignum}
\epsscale{1.00}
\plotone{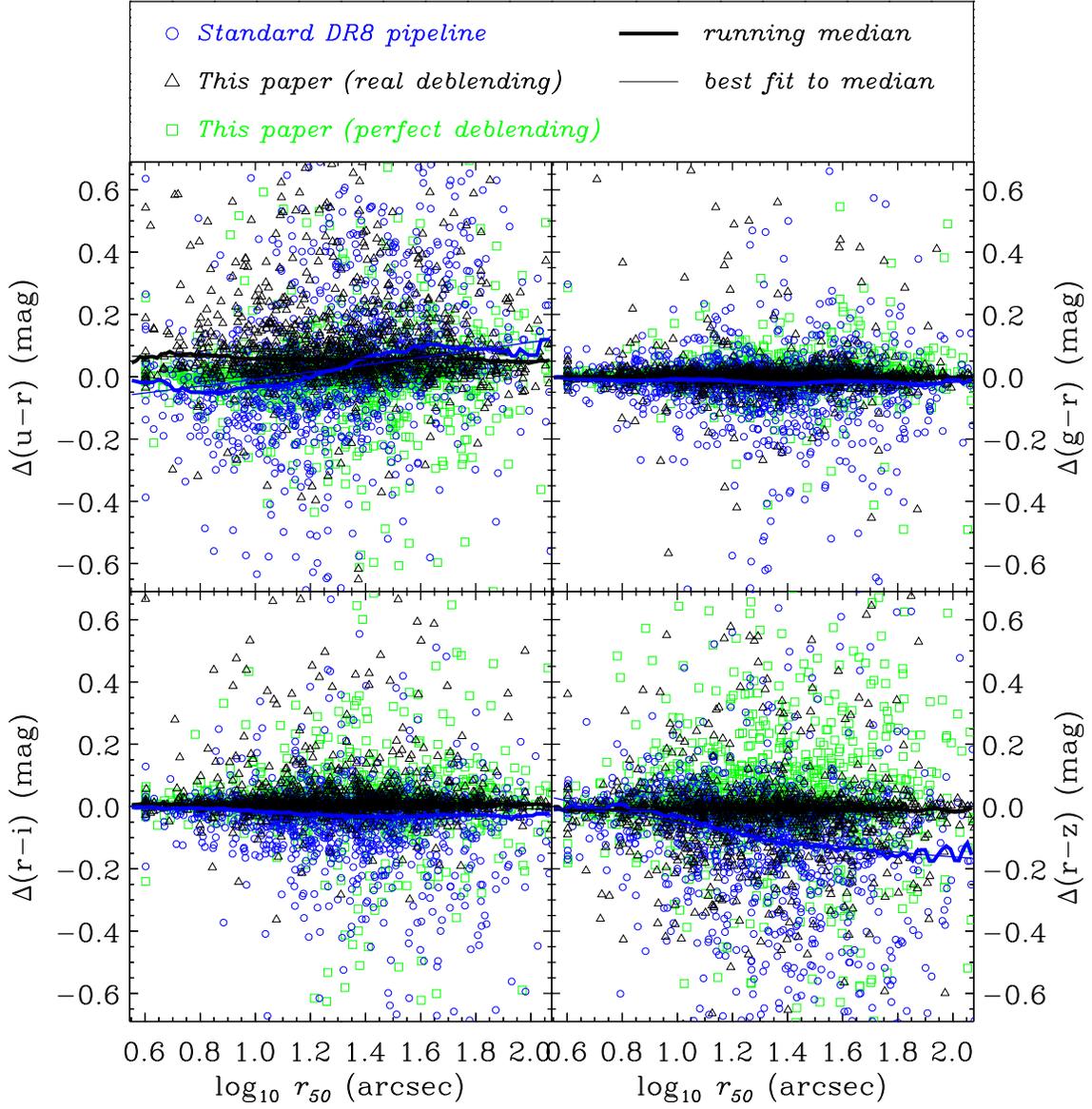}
\caption{\label{fig:sky_offsets_ugiz} Similar to Figure
  \ref{fig:sky_offsets_ronly}, but showing errors in colors with respect to
  $r$, using the $u$, $g$, $i$ and $z$ bands.}
\end{figure}

To test more realistic scenarios, Figures \ref{fig:sky_offsets_ronly}
and \ref{fig:sky_offsets_ugiz} also show results from an automated
analysis of these mosaics, performed without knowledge of the true
image or of the original image without the fake galaxy. We will
compare these results to the standard SDSS catalog analysis.

We have tested two versions of {\tt photo}: the one used in the DR7
catalog ({\tt v5\_4}) and the one used in the DR8 catalog ({\tt
  v5\_6}). The {\tt v5\_6} version was designed to perform sky
subtraction better than {\tt v5\_4}.  As found by \citet{aihara11a},
in practice {\tt v5\_6} is a minor improvement, but the differences
between the two pipelines are small.

In these figures, the circles are the SDSS {\tt photo} {\tt v5\_6}
measurements of the fake galaxies, using the cModel magnitude.  The
top panel of Figure \ref{fig:sky_offsets_ronly} shows the error in the
galaxy half-light radius in the $r$-band. There is a large
underestimation of the fluxes and radii --- it becomes signficant at
$r_{50}>10$ arcsec. The thick solid line running through the points is
a running median estimate of the residual (inside a window of $\pm
0.15$ dex). The thin solid line is a quadratic approximation to the
running median of the form:
\begin{equation}
\label{eq:sky_offsets_model}
a_0 + a_1 \log_{10}\left(r_{50}/10''\right)
+ a_2 \left[\log_{10}\left(r_{50}/10''\right)\right]^2
\end{equation}
Table \ref{table:sky_offsets} lists the parameters of this fit as well
as the standard deviation $\sigma_a$ around the median for both
versions of {\tt photo}.

The errors in color shown in Figure \ref{fig:sky_offsets_ugiz} are
much smaller, though there are detectable trends in $u$ and $z$. Table
\ref{table:sky_offsets} also lists the fits to the color residuals.

We also investigated these residuals as a function of galaxy area
(that is, $\pi r_{50}^2 (b/a)$, where $b/a$ is the axis ratio), as
advocated by \citet{west10a}. There was little substantial reduction
in the scatter in these residuals, and qualitatively the results were
very similar.

We emphasize that for the primary SDSS spectroscopic sample at $z\sim
0.1$ the issue is not nearly as dire as it appears in this plot ---
only the largest such galaxies have $r_{50}\sim 10$--20 arcsec, for
which the effects are still relatively small.

For comparison, we have performed deblending and photometry on our sky
subtracted images.  Our algorithms are similar to those used in {\tt
  photo}, but are tuned to work for large galaxies at the expense of
small galaxies. Essentially, we detect pixels 5$\sigma$ above
background and connect those pixels into ``parent'' objects. We find
peaks within each parent and remove those that are consistent with the
image point-spread function. We then smooth the remaining image using
a Gaussian with $\sigma=4$ pixels and redetect peaks: this step will
retain large galaxies without breaking them up into many children, but
essentially ignore small ones. Then we define a symmetric template
around each galaxy peak and extract the flux in each pixel associated
with each peak in the same manner that {\tt photo} does. To make the
procedure more robust, we detect galaxies and define templates only in
the $r$-band, and then use those same templates for all
bands. Finally, we fit a two-dimensional \Sersic\ profile to the
deblended galaxy image to estimate its half-light radius and flux in
each band.

The triangles in Figures \ref{fig:sky_offsets_ronly} and
\ref{fig:sky_offsets_ugiz} show the results of this procedure. Again,
the lines show the running median and the quadratic fit to it (with
parameters listed in Table \ref{table:sky_offsets} in the row marked
``global''). This analysis performs substantially better than {\tt
  photo}: the size errors exist but are much smaller, while the flux
errors are about 0.1 mag independent of size. In addition, the scatter
$\sigma_a$ is about a factor of two smaller, which probably results
from a smaller dependence of the errors on the \Sersic\ index and axis
ratio of the galaxies.

\subsection{Comparison to previous results}
\label{sec:comparison}

\begin{figure}
\figurenum{\fignum}
\epsscale{1.00}
\plotone{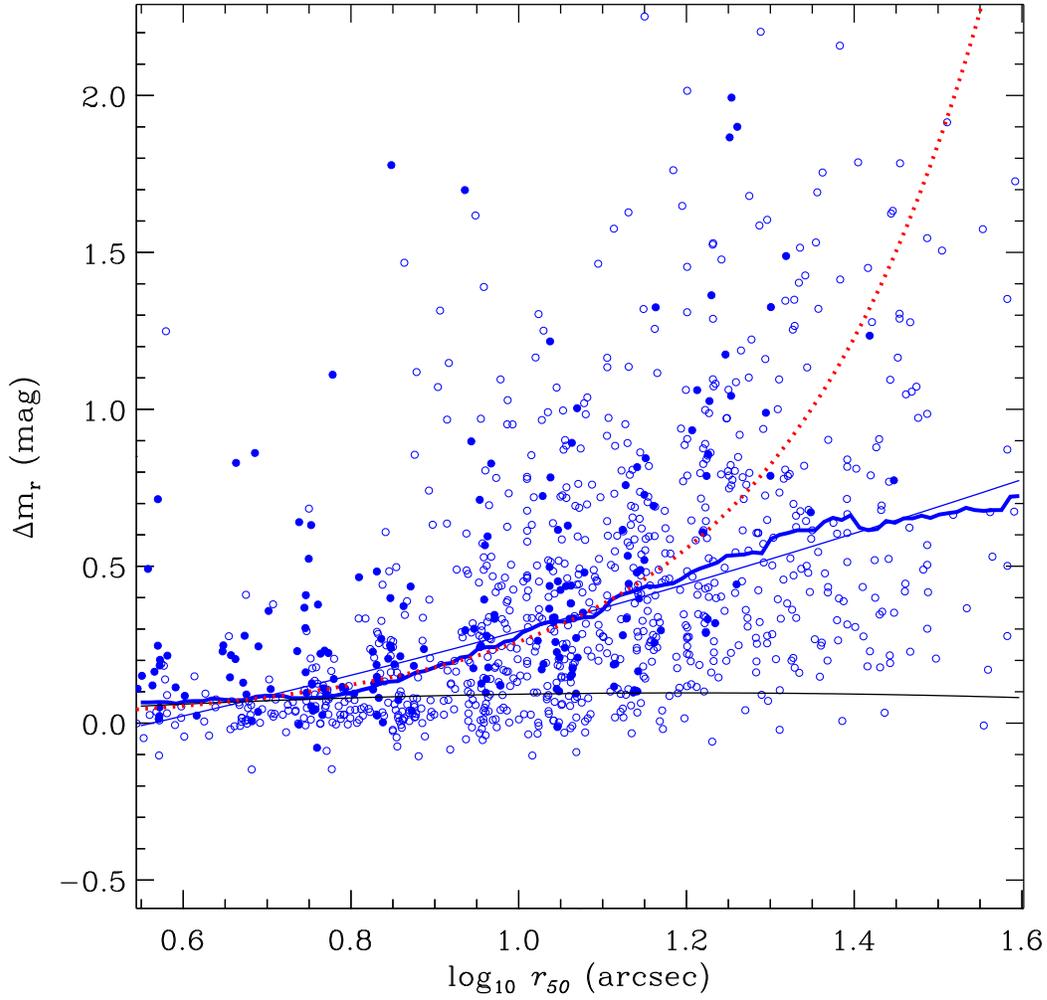}
\caption{\label{fig:sky_offsets_vs_r50meas} Similar to Figure
  \ref{fig:sky_offsets_ronly}, but showing errors $r$-band magnitude against
  the measured value of $r_{50}$ rather than the true value. Open
  circles show galaxies with \Sersic\ indices $n< 3.5$, and filled
  circles show galaxies with $n>3.5$. The
  dotted line shows the functional form given by \citet{hyde09a}.}
\end{figure}

\citet{hyde09a} found similar results to what we find here, but for
elliptical galaxies. For comparison to their results, Figure
\ref{fig:sky_offsets_vs_r50meas} shows the magnitude errors of {\tt
  photo} {\tt v5\_6} in the $r$-band, as a function of the measured
$r_{50}$ (not the true value).  Again, the lines show the running
median and the fit. The dotted line shows the parametrization of the
error given by \citet{hyde09a}. At small sizes it is in good agreement
with our results, but at large sizes it exceeds our corrections.
Almost certainly this is because our result averages over all galaxy
profiles. The solid points in Figure \ref{fig:sky_offsets_vs_r50meas}
indicate fake galaxies with \Sersic\ index $n>3.5$, which follow much
better the results of \citet{hyde09a}.

\citet{west10a} also explored these issues, but for a broader
distribution of galaxy types.  They published a fitting function based
on the SDSS catalog axis ratios, 90\% light radii, and magnitudes
(their Equation 1). They predict the flux ``lost'' in the standard
reductions $f_{\mathrm{lost}} = f_{\mathrm{meas}} - f_{\mathrm{true}}$
(in nanomaggies). In this context, both $f_{\mathrm{meas}}$ and
$f_{\mathrm{true}}$ are standard SDSS Petrosian fluxes
(\citealt{blanton01a, stoughton02a}).  \citet{west10a} suggest the
simple form:
\begin{equation}
\label{eq:west}
\log_{10} f_{\mathrm{lost}} = 
b_0 + 
b_1 \log_{10} m_r + 
b_2 \log_{10} r_{90} + 
b_3 \log_{10} b/a,
\end{equation}
where $m_r$, $r_{90}$ and $b/a$ are the Petrosian quantities reported
by the SDSS catalog. We have refit the parameters of this model to our
simulated data.  Table \ref{table:west} lists the values of the
parameters of this fit, for both versions of {\tt photo} studied here,
and evaluating the flux lost in both the Petrosian and cModel
estimates.

\begin{figure}
\figurenum{\fignum}
\epsscale{1.00}
\plotone{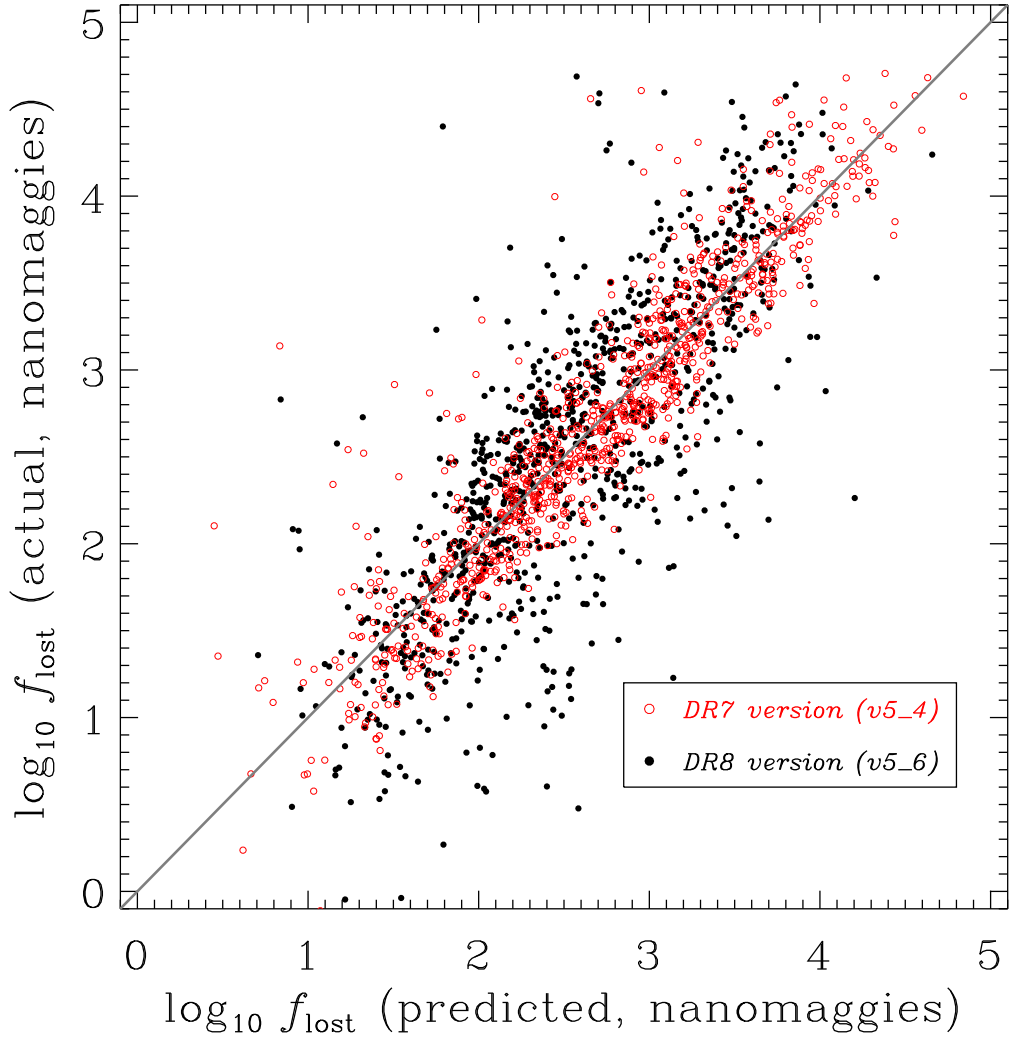}
\caption{\label{fig:sky_offsets_vs_west} Actual lost flux compared to
  that predicted using the functional form in Equation \ref{eq:west}
  (proposed by \citealt{west10a}), with the coefficients from Table
  \ref{eq:west}.  We show the actual flux lost (in nanomaggies) in the
  standard SDSS measurements relative to the total Petrosian flux of
  each fake galaxy, against a prediction based on the measured
  Petrosian magnitude, 90\% light radius and axis ratio $b/a$. The
  filled circles show the results for the DR8 version of the pipeline,
  and the open circles show the results for the DR7 version of the
  pipeline.}
\end{figure}

We demonstrate the quality of fit for this formula in Figure
\ref{fig:sky_offsets_vs_west}, showing the actual flux lost relative
to the predicted value. The open circles show results for the DR7
version of the photometry ({\tt photo v5\_4}) whereas the filled
circles show results for the DR8 version of the photometry ({\tt photo
  v5\_6}).  The predictions perform excellently on DR7 data.  However,
for DR8 data there is greater scatter.  We find a correlation between
the scatter and $m_r$ that the simple function above fails to explain;
in particular, the outliers below the relation are typically galaxies
with $m_r<15$.  These differences are likely due to the DR8 photometry
being slightly better in the large size, high flux regime
(\citealt{aihara11a}).

The coefficients for Equation \ref{eq:west} derived by \citet{west10a}
are not directly comparable to those in Table \ref{table:west}. Their
comparison only measured the flux lost due to sky subtraction alone,
and did not include the effects of {\tt photo}'s deblending. Our
comparison is to the {\tt photo} catalog itself and therefore includes
the effects of deblending.  It turns out that in the SDSS pipeline
results, a substantial function of the lost flux is due to effects of
poor deblending. Therefore, the coefficients found here would be most
applicable to an actual SDSS photometric catalog sample.

However, we do not necessarily recommend using these formulae for
robust determinations of fluxes for bright and/or large galaxies ---
reanalyzing properly background subtracted images is clearly a better
route.  However, they are excellent ways to determine when such
effects are important in the SDSS catalog.

\subsection{Direct comparison to Montage} 
\label{sec:galmontage}

As another test, we have also taken a sample of nearby bright galaxies
and compared large-aperture photometry between our images and images
from Montage. We randomly select 200 galaxies brighter than $r=15$ to
perform this test. We compare aperture photometry within a radius of
$1.3 r_{90}$ for each galaxy. In general, the fluxes and colors are in
good agreement. The median ratio of the flux estimated from the
Montage images to our flux estimate to is about 96\%, as for the point
source estimates above (and likely for the same reason). The standard
deviations around this mean for the $(u,g,r,i,z)$ bands are
$(5,2,1,2,3)\%$, indicating very good agreement between the methods.

\section{ Summary}
\label{sec:summary}

In this paper, we have presented a method to estimate the sky
background in SDSS images, suitable the purposes of measuring large,
bright galaxies.  While this model has no physical motivation, it
appears to have the right balance of freedom and physical constraints
to adequately describe most of the SDSS imaging.

We have tested the measurements of point sources in the sky subtracted
images, to ensure that they are not signficantly degraded with respect
to the original SDSS data.  In addition, we have inserted fake
galaxies into the data and reran the sky subtraction to evaluate how
well our procedure works for bright galaxies.  We find substantial
improvement with respect to the standard pipelines.  Whereas for the
standard SDSS {\tt photo} pipeline the sky subtraction becomes
important at the 20\% level at $r_{50}>10$ arcsec or so, we do not see
comparable errors in our results even up to $r_{50} \sim 100$ arcsec.

We have compared our results to a similar service provided by IRSA,
the Montage package. Essentially, our results are similar to theirs
for large objects, though we identify a slight error in their overall
fluxes. For bright stars their interpolation method appears to
introduce some very slight errors.

These results are now publicly available as part of the SDSS-III DR8
release. Users can acquire individual sky subtracted and calibrated
fields, as well as request large-scale mosaics in FITS format.

\acknowledgments

We thank Richard Cool, David Hogg, Rachel Mandelbaum, Michael Strauss,
and particularly Andrew A.~West for discussions of the methods and
results found here.

The authors acknowledge funding support from NASA grants
06-GALEX06-0030, NNX09AC85G and NNX09AC95G, and \emph{Spitzer} grant
G05-AR-50443, as well as a Google Research Award. This research has
made use of NASA's Astrophysics Data System and of the NASA/IPAC
Extragalactic Database (NED) which is operated by the Jet Propulsion
Laboratory, California Institute of Technology, under contract with
the National Aeronautics and Space Administration.

Funding for SDSS-III has been provided by the Alfred P. Sloan
Foundation, the Participating Institutions, the National Science
Foundation, and the U.S. Department of Energy. The SDSS-III web site
is http://www.sdss3.org/.

SDSS-III is managed by the Astrophysical Research Consortium for the
Participating Institutions of the SDSS-III Collaboration including the
University of Arizona, the Brazilian Participation Group, Brookhaven
National Laboratory, University of Cambridge, University of Florida,
the French Participation Group, the German Participation Group, the
Instituto de Astrofisica de Canarias, the Michigan State/Notre
Dame/JINA Participation Group, Johns Hopkins University, Lawrence
Berkeley National Laboratory, Max Planck Institute for Astrophysics,
New Mexico State University, New York University, the Ohio State
University, University of Portsmouth, Princeton University, University
of Tokyo, the University of Utah, Vanderbilt University, University of
Virginia, University of Washington, and Yale University.

This research made use of Montage, funded by the National Aeronautics and
Space Administration's Earth Science Technology Office, Computation
Technologies Project, under Cooperative Agreement Number NCC5-626 between
NASA and the California Institute of Technology. Montage is maintained by
the NASA/IPAC Infrared Science Archive.

\newpage
\clearpage
\clearpage

\setcounter{thefigs}{0}

\clearpage
\stepcounter{thefigs}

\clearpage
\stepcounter{thefigs}

\clearpage
\stepcounter{thefigs}

\clearpage
\stepcounter{thefigs}

\clearpage
\stepcounter{thefigs}

\clearpage
\stepcounter{thefigs}

\clearpage
\stepcounter{thefigs}

\clearpage
\stepcounter{thefigs}

\clearpage
\stepcounter{thefigs}

\clearpage
\stepcounter{thefigs}

\clearpage
\stepcounter{thefigs}

\clearpage
\stepcounter{thefigs}

\clearpage
\stepcounter{thefigs}

\clearpage
\stepcounter{thefigs}

\clearpage
\stepcounter{thefigs}

\newpage
\clearpage
\clearpage

\setcounter{thetabs}{0}

\clearpage

\stepcounter{thetabs}
\begin{deluxetable}{llllll}
\tablewidth{0pt} \tablecolumns{6} \tablenum{\tabnum}
\tablecaption{\label{table:sky_offsets} Fits to residual sizes and
  fluxes} \tablehead{ Quantity & Sky version & $a_0$ & $a_1$ & $a_2$ &
  $\sigma_a$ } 
\startdata 
$r_{50}$ & \texttt{v5\_4} & $-0.151$ & $-0.211$ & $-0.262$ & $0.134$\cr
 & \texttt{v5\_6} & $-0.146$ & $-0.193$ & $-0.269$ & $0.129$\cr
 & global & $-0.036$ & $0.034$ & $-0.215$ & $0.086$\cr
$m_r$ & \texttt{v5\_4} & $0.168$ & $0.641$ & $0.704$ & $0.173$\cr
 & \texttt{v5\_6} & $0.101$ & $0.486$ & $0.861$ & $0.237$\cr
 & global & $0.079$ & $0.032$ & $-0.001$ & $0.163$\cr
$(u-r)$ & \texttt{v5\_4} & $0.028$ & $0.184$ & $-0.047$ & $0.147$\cr
 & \texttt{v5\_6} & $-0.008$ & $0.113$ & $0.006$ & $0.198$\cr
 & global & $0.062$ & $-0.009$ & $-0.001$ & $0.073$\cr
$(g-r)$ & \texttt{v5\_4} & $-0.008$ & $-0.007$ & $0.004$ & $0.045$\cr
 & \texttt{v5\_6} & $-0.013$ & $-0.015$ & $0.010$ & $0.066$\cr
 & global & $0.001$ & $0.002$ & $-0.005$ & $0.016$\cr
$(r-i)$ & \texttt{v5\_4} & $-0.017$ & $-0.025$ & $0.040$ & $0.056$\cr
 & \texttt{v5\_6} & $-0.018$ & $-0.036$ & $0.026$ & $0.074$\cr
 & global & $0.008$ & $0.002$ & $-0.003$ & $0.021$\cr
$(r-z)$ & \texttt{v5\_4} & $-0.054$ & $-0.129$ & $0.122$ & $0.142$\cr
 & \texttt{v5\_6} & $-0.051$ & $-0.156$ & $0.046$ & $0.165$\cr
 & global & $-0.011$ & $-0.001$ & $0.001$ & $0.053$\cr
$m_r$ (vs $r_{50}$ meas.) & \texttt{v5\_4} & $0.379$ & $0.959$ & $0.530$ & $0.325$\cr
 & \texttt{v5\_6} & $0.293$ & $0.727$ & $0.131$ & $0.290$\cr
 & global & $0.092$ & $0.040$ & $-0.098$ & $0.177$\cr
\enddata
\tablecomments{ Listed parameters refer to Equation
  \ref{eq:sky_offsets_model}. Magnitude offsets are in units of
  magnitudes.  Half-light radius offsets are in units of dex. Results
  labeled {\tt v5\_4} are appropriate for the SDSS DR7
  catalog. Results labeled {\tt v5\_6} are appropriate for SDSS DR8
  catalog. Results labeled ``global'' are appropriate for the
  sky-subtraction and deblending analysis described in this
  paper. These fits are appropriate between a true $r_{50}$ of 5 and
  100 arcsec, or in the case of the last three rows a measured
  $r_{50}$ of 5 and 40 arcsec. See \S\ref{sec:endtoend} for more
  information.}
\end{deluxetable}

\stepcounter{thetabs}
\begin{deluxetable}{lllllll}
\tablewidth{0pt} \tablecolumns{7} \tablenum{\tabnum}
\tablecaption{\label{table:west} Multiparameter fits to lost flux}
\tablehead{ Flux type & {\tt photo} version & $b_0$ & $b_1$ & $b_2$ &
  $b_3$ & $\sigma_b$ } 
\startdata 
Petrosian & {\tt v5\_4}  & 6.359 & -6.609 & 3.129 & 0.767 & 0.291 \cr
 & {\tt v5\_6}  & 4.643 & -4.487 & 2.320 & 0.614 & 0.540 \cr
cModel & {\tt v5\_4}  & 11.784 & -10.470 & 2.453 & 0.538 & 0.357 \cr
 & {\tt v5\_6}  & 9.510 & -8.239 & 1.995 & 0.362 & 0.509 \cr
\enddata
\tablecomments{ The listed parameters refer to Equation
  \ref{eq:west}. $\sigma_b$ lists r.m.s.~deviation around the model in
  dex. }
\end{deluxetable}


\begin{thebibliography}{31}
\expandafter\ifx\csname natexlab\endcsname\relax\def\natexlab#1{#1}\fi

\bibitem[{{Abazajian} {et~al.}(2009){Abazajian}, {Adelman-McCarthy},
  {Ag{\"u}eros}, {Allam}, {Allende Prieto}, {An}, {Anderson}, {Anderson},
  {Annis}, {Bahcall}, \& et~al.}]{abazajian09a}
{Abazajian}, K.~N., {Adelman-McCarthy}, J.~K., {Ag{\"u}eros}, M.~A., {Allam},
  S.~S., {Allende Prieto}, C., {An}, D., {Anderson}, K.~S.~J., {Anderson},
  S.~F., {Annis}, J., {Bahcall}, N.~A., \& et~al. 2009, {The Seventh Data
  Release of the Sloan Digital Sky Survey}

\bibitem[{{Aihara} {et~al.}(2011)}]{aihara11a}
{Aihara}, H. {et~al.} 2011, ArXiv e-prints

\bibitem[{{Bernardi} {et~al.}(2007){Bernardi}, {Hyde}, {Sheth}, {Miller}, \&
  {Nichol}}]{bernardi07a}
{Bernardi}, M., {Hyde}, J.~B., {Sheth}, R.~K., {Miller}, C.~J., \& {Nichol},
  R.~C. 2007, \aj, 133, 1741

\bibitem[{{Berriman} {et~al.}(2004){Berriman}, {Deelman}, {Good}, {Jacob},
  {Katz}, {Kesselman}, {Laity}, {Prince}, {Singh}, \& {Su}}]{berriman04a}
{Berriman}, G.~B., {Deelman}, E., {Good}, J.~C., {Jacob}, J.~C., {Katz}, D.~S.,
  {Kesselman}, C., {Laity}, A.~C., {Prince}, T.~A., {Singh}, G., \& {Su}, M.-H.
  2004, in Society of Photo-Optical Instrumentation Engineers (SPIE) Conference
  Series, Vol. 5493, Society of Photo-Optical Instrumentation Engineers (SPIE)
  Conference Series, ed. P.~J. {Quinn} \& A.~{Bridger}, 221--232

\bibitem[{{Berriman} {et~al.}(2003){Berriman}, {Good}, {Curkendall}, {Jacob},
  {Katz}, {Prince}, \& {Williams}}]{berriman03a}
{Berriman}, G.~B., {Good}, J.~C., {Curkendall}, D.~W., {Jacob}, J.~C., {Katz},
  D.~S., {Prince}, T.~A., \& {Williams}, R. 2003, in Astronomical Society of
  the Pacific Conference Series, Vol. 295, Astronomical Data Analysis Software
  and Systems XII, ed. H.~E. {Payne}, R.~I. {Jedrzejewski}, \& R.~N. {Hook},
  343--346

\bibitem[{{Blanton} {et~al.}(2005{\natexlab{a}}){Blanton}, {Eisenstein},
  {Hogg}, {Schlegel}, \& {Brinkmann}}]{blanton05b}
{Blanton}, M.~R., {Eisenstein}, D., {Hogg}, D.~W., {Schlegel}, D.~J., \&
  {Brinkmann}, J. 2005{\natexlab{a}}, \apj, 629, 143

\bibitem[{{Blanton} {et~al.}(2005{\natexlab{b}}){Blanton}, {Lupton},
  {Schlegel}, {Strauss}, {Brinkmann}, {Fukugita}, \& {Loveday}}]{blanton04b}
{Blanton}, M.~R., {Lupton}, R.~H., {Schlegel}, D.~J., {Strauss}, M.~A.,
  {Brinkmann}, J., {Fukugita}, M., \& {Loveday}, J. 2005{\natexlab{b}}, \apj,
  631, 208

\bibitem[{{Blanton} {et~al.}(2001)}]{blanton01a}
{Blanton}, M.~R. {et~al.} 2001, \aj, 121, 2358

\bibitem[{{Blanton} {et~al.}(2005{\natexlab{c}})}]{blanton05a}
{Blanton}, M.~R. {et~al.} 2005{\natexlab{c}}, \aj, 129, 2562

\bibitem[{{de Vaucouleurs} {et~al.}(1991){de Vaucouleurs}, {de Vaucouleurs},
  {Corwin}, {Buta}, {Paturel}, \& {Fouque}}]{devaucouleurs91a}
{de Vaucouleurs}, G., {de Vaucouleurs}, A., {Corwin}, Jr., H.~G., {Buta},
  R.~J., {Paturel}, G., \& {Fouque}, P. 1991, {Third Reference Catalogue of
  Bright Galaxies} (Berlin: Springer-Verlag)

\bibitem[{{Eisenstein} {et~al.}(2011){Eisenstein}, {Weinberg}, {Agol},
  {Aihara}, {Allende Prieto}, {Anderson}, {Arns}, {Aubourg}, {Bailey},
  {Balbinot}, \& et~al.}]{eisenstein11a}
{Eisenstein}, D.~J., {Weinberg}, D.~H., {Agol}, E., {Aihara}, H., {Allende
  Prieto}, C., {Anderson}, S.~F., {Arns}, J.~A., {Aubourg}, E., {Bailey}, S.,
  {Balbinot}, E., \& et~al. 2011, ArXiv e-prints

\bibitem[{Fukugita {et~al.}(1996)Fukugita, Ichikawa, Gunn, Doi, Shimasaku, \&
  Schneider}]{fukugita96a}
Fukugita, M., Ichikawa, T., Gunn, J.~E., Doi, M., Shimasaku, K., \& Schneider,
  D.~P. 1996, \aj, 111, 1748

\bibitem[{{Greisen} \& {Calabretta}(2002)}]{greisen02a}
{Greisen}, E.~W. \& {Calabretta}, M.~R. 2002, \aap, 395, 1061

\bibitem[{{Gunn} {et~al.}(2006)}]{gunn05a}
{Gunn}, J.~E. {et~al.} 2006, \aj, 131, 2332

\bibitem[{{Hogg} {et~al.}(2001){Hogg}, {Finkbeiner}, {Schlegel}, \&
  {Gunn}}]{hogg01a}
{Hogg}, D.~W., {Finkbeiner}, D.~P., {Schlegel}, D.~J., \& {Gunn}, J.~E. 2001,
  \aj, 122, 2129

\bibitem[{{Hyde} \& {Bernardi}(2009)}]{hyde09a}
{Hyde}, J.~B. \& {Bernardi}, M. 2009, \mnras, 394, 1978

\bibitem[{{Lauer} {et~al.}(2007)}]{lauer07a}
{Lauer}, T.~R. {et~al.} 2007, \apj, 662, 808

\bibitem[{{Lupton} {et~al.}(2001){Lupton}, {Gunn}, {Ivezi{\'c}}, {Knapp}, \&
  {Kent}}]{lupton01a}
{Lupton}, R., {Gunn}, J.~E., {Ivezi{\'c}}, {\v Z}., {Knapp}, G.~R., \& {Kent},
  S. 2001, in Astronomical Society of the Pacific Conference Series, Vol. 238,
  Astronomical Data Analysis Software and Systems X, ed. F.~R. {Harnden}, Jr.,
  F.~A. {Primini}, \& H.~E. {Payne}, 269--278

\bibitem[{{Mandelbaum} {et~al.}(2006){Mandelbaum}, {Seljak}, {Cool}, {Blanton},
  {Hirata}, \& {Brinkmann}}]{mandelbaum06a}
{Mandelbaum}, R., {Seljak}, U., {Cool}, R.~J., {Blanton}, M., {Hirata}, C.~M.,
  \& {Brinkmann}, J. 2006, \mnras, 372, 758

\bibitem[{{Masjedi} {et~al.}(2006)}]{masjedi06a}
{Masjedi}, M. {et~al.} 2006, \apj, 644, 54

\bibitem[{{Padmanabhan} {et~al.}(2007)}]{padmanabhan07b}
{Padmanabhan}, N. {et~al.} 2007, \mnras, 378, 852

\bibitem[{{Padmanabhan} {et~al.}(2008)}]{padmanabhan07a}
{Padmanabhan}, N. {et~al.} 2008, \apj, 674, 1217

\bibitem[{{Park} \& {Schowengerdt}(1983)}]{park83a}
{Park}, S.~K. \& {Schowengerdt}, R.~A. 1983, Computer Graphics Image
  Processing, 23, 258

\bibitem[{Pier {et~al.}(2003)Pier, Munn, Hindsley, Hennessy, Kent, Lupton, \&
  {Ivezi{\' c}}}]{pier03a}
Pier, J.~R., Munn, J.~A., Hindsley, R.~B., Hennessy, G.~S., Kent, S.~M.,
  Lupton, R.~H., \& {Ivezi{\' c}}, {\v Z}. 2003, \aj, 125, 1559

\bibitem[{{Price-Whelan} \& {Hogg}(2010)}]{pricewhelan10a}
{Price-Whelan}, A.~M. \& {Hogg}, D.~W. 2010, \pasp, 122, 207

\bibitem[{{Smith} {et~al.}(2002)}]{smith02a}
{Smith}, J.~A. {et~al.} 2002, \aj, 123, 2121

\bibitem[{Stoughton {et~al.}(2002)}]{stoughton02a}
Stoughton, C. {et~al.} 2002, \aj, 123, 485

\bibitem[{{West}(2005)}]{west05a}
{West}, A.~A. 2005, Ph.D.~Thesis

\bibitem[{{West} {et~al.}(2010){West}, {Garcia-Appadoo}, {Dalcanton}, {Disney},
  {Rockosi}, {Ivezi{\'c}}, {Bentz}, \& {Brinkmann}}]{west10a}
{West}, A.~A., {Garcia-Appadoo}, D.~A., {Dalcanton}, J.~J., {Disney}, M.~J.,
  {Rockosi}, C.~M., {Ivezi{\'c}}, {\v Z}., {Bentz}, M.~C., \& {Brinkmann}, J.
  2010, \aj, 139, 315

\bibitem[{{York} {et~al.}(2000)}]{york00a}
{York}, D.~G. {et~al.} 2000, \aj, 120, 1579

\bibitem[{{Zhu} {et~al.}(2010){Zhu}, {Blanton}, \& {Moustakas}}]{zhu10a}
{Zhu}, G., {Blanton}, M.~R., \& {Moustakas}, J. 2010, \apj, 722, 491

\end{thebibliography}
\end{document}